\documentclass[sn-mathphys,Numbered]{sn-jnl}

\usepackage{graphicx}%
\usepackage{multirow}%
\usepackage{amsmath,amssymb,amsfonts}%
\usepackage{amsthm}%
\usepackage{mathrsfs}%
\usepackage[title]{appendix}%
\usepackage{xcolor}%
\usepackage{textcomp}%
\usepackage{manyfoot}%
\usepackage{booktabs}%
\usepackage{algorithm}%
\usepackage{algorithmicx}%
\usepackage{algpseudocode}%
\usepackage{listings}%
\usepackage{subcaption}
\usepackage{csquotes}

\theoremstyle{thmstyleone}%
\theoremstyle{thmstyletwo}%

\theoremstyle{thmstylethree}%

\begin{document}

\title[Article Title]{An Exploratory Analysis of COVID Bot vs Human Disinformation Dissemination stemming from the Disinformation Dozen on Telegram}

\author*[1]{Lynnette Hui Xian Ng}\email{huixiann@andrew.cmu.edu}

\author[1]{Ian Kloo}\email{ipk@andrew.cmu.edu}

\author[1]{Samantha Clark}\email{samanthc@andrew.cmu.edu}

\author[1]{Kathleen M. Carley}\email{carley@andrew.cmu.edu}

\affil*[1]{\orgdiv{Center for Informed Democracy \& Social - Cybersecurity (IDeaS)}, \orgname{Carnegie Mellon University}, \orgaddress{\street{5000 Forbes Ave}, \city{Pittsburgh}, \postcode{15213}, \state{Pennsylvania}, \country{USA}}}

\abstract{The COVID-19 pandemic of 2021 led to a worldwide health crisis that was accompanied by an infodemic. A group of 12 social media personalities, dubbed the ``Disinformation Dozen", were identified as key in spreading disinformation regarding the COVID-19 virus, treatments, and vaccines. This study focuses on the spread of disinformation propagated by this group on Telegram, a mobile messaging and social media platform. After segregating users into three groups -- the Disinformation Dozen, bots, and humans --, we perform an investigation with a dataset of Telegram messages from January to June 2023, comparatively analyzing temporal, topical, and network features. We observe that the Disinformation Dozen are highly involved in the initial dissemination of disinformation but are not the main drivers of the propagation of disinformation. Bot users are extremely active in conversation threads, while human users are active propagators of information, disseminating posts between Telegram channels through the forwarding mechanism. }


\maketitle

\section{Introduction}
\label{sec:intro}
The Coronavirus Pandemic of 2020 led to a worldwide health crisis. In a world physically interconnected via air, land, and sea travel, the virus spread across international borders exceptionally quickly. To minimize human-to-human contact and mitigate the spread of the virus, governments closed borders and enforced lockdowns where people were only allowed to leave their homes for essential tasks. In late 2020, several pharmaceutical companies developed vaccines that were effective at preventing and reducing the symptoms of COVID-19 infections, and in December 2020, the worldwide vaccination campaign began.

As the virus spread, so did many mis/disinformation narratives. These included theories that COVID-19 was a bioweapon built for the destruction of mankind, that the 5G mobile network caused COVID-19, or that the threat posed by COVID-19 was exaggerated \cite{ng2021coronavirus,uscinski2020people}. Research has shown that the foundations of these disinformation narratives are psychological predispositions to reject information coming from authority figures and a propensity to view major events as a product of conspiracy, partisan, and ideological motivations \cite{uscinski2020people}. The large amount of disinformation spread during the COVID-19 pandemic became known as an ''infodemic." 

After investigating the sources of COVID-19 disinformation, the Center for Countering Digital Hate (CCDH) identified twelve people as the most responsible for propagating these narratives. They dubbed this group the ``Disinformation Dozen" \cite{counterhateDisinformationDozen}. From over 800,000 posts extracted from Facebook and Twitter in early 2021, the CCDH reported that at least 65\% of anti-vaccine content could be attributed to the Disinformation Dozen. These users were highly visible, with millions of followers, and were termed ``repeat offenders" of disinformation spread. 

The report produced by CCDH analyzed the influence of the Disinformation Dozen on traditional social media platforms like Facebook, Twitter, and Instagram, observing that they had high influence because of their voluminous posts and enormous following. In terms of information propagation, two main groups of users have been observed to share the content generated by the Disinformation Dozen on Twitter: low-credibility media outlets and politically-linked accounts \cite{nogara2022disinformation}. In this study, we extend the investigation of information dissemination from social media platforms to mobile media platforms, in particular, the Telegram messaging application. 

While the Disinformation Dozen's activities were initially identified on Facebook, Twitter, and Instagram, these platforms each made efforts to combat COVID-19 disinformation during the pandemic. This censorship ranged from flagging potentially misleading posts to platform bans for certain accounts \cite{krishnan2021research}. While some of these interventions were effective, they also served to push spreaders of disinformation onto other platforms.  Telegram is a messaging service with social media functionality that only censors content and users in the most extreme situations (e.g., removing content that is explicitly illegal).  We found that 8 of the 12 Disinformation Dozen were still active on Telegram in early 2023, so we sought to investigate the disinformation networks surrounding these accounts on the platform.

Telegram has become a popular mobile messaging platform, with at least 700 million monthly active users worldwide \cite{forbesPavelDurov}. Its large user base provides ripe ground for COVID-19 discourse. Emerging disinformation stories on Telegram have been found by identifying skepticism of claims and co-occurrence with colloquial uses of government acts enacted to combat online falsehoods \cite{ng2020analyzing}. Many Telegram channels have been discovered to be aggregating conspiracy theories surrounding COVID-19, evolving these theories as the channels develop, and spreading distrust in governments \cite{willaert2022disinformation}. With the large number of coronavirus-related information circulating on Telegram, the messaging app is a huge source of disinformation transmission \cite{sosa2022multimodal}, and therefore we leverage this wealth of information in our study.

In this paper, we investigated the disinformation spread by the Disinformation Dozen on the Telegram messaging application. From the Telegram channels hosted by the Disinformation Dozen, we collected channel messages and information about forwarded and replied messages. We analyzed the temporal, topic, and network interactions among three different sets of users: the Disinformation Dozen, bot users, and regular human users, generating insights into the nature of information spread through Telegram interactions. Our key findings suggest that while the Disinformation Dozen may be the initial seed in setting up Telegram channels and initiating information spread, bots are extremely active in generating engagement through messaging replies, while humans facilitate the spread of disinformation to other channels through the forwarding mechanic. 

\section{Research Questions and Contributions}
\label{sec:rq}
We ask the following Research Questions with respect to the dissemination of disinformation that stemmed from the Disinformation Dozen on Telegram: 
\begin{enumerate}
    \item RQ1: How do the temporal trends of message posting patterns differ from the Disinformation Dozen, Bots, and Human users? 
    \item RQ2: How do the topics posted within messages differ for the Disinformation Dozen, Bots, and Human users?
    \item RQ3: How does disinformation begin spreading from the Disinformation Dozen throughout the broader Telegram network?
\end{enumerate}

Drawing data from the Telegram messaging application, we analyzed the activity of disinformation dissemination that stemmed from the Disinformation Dozen. 

The main contributions of our work can be summarized as follows: 
\begin{enumerate}
    \item We gathered a novel dataset that encompasses the messages published on Telegram by the Disinformation Dozen, the messages within the channels that they control, and the messages of the channels from which the original messages were forwarded.
    \item We investigated the patterns of information dissemination within the Telegram messenger application through temporal, topical, and network analysis. We uncovered distinct traits in terms of messaging patterns and dissemination across the Disinformation Dozen, bot users, and human users on Telegram.
    \item We adapted a current bot detection algorithm for Telegram, designing and validating a Telegram bot detection algorithm that works with an accuracy of 72\% through manual verification. We also summarized the characteristics of bot-like users through our observations from the manual annotation of Telegram users.
\end{enumerate}

\subsection{Structure of the Paper}
The structure of the paper is as follows. Having established an introduction and our research questions in \autoref{sec:intro} and \autoref{sec:rq}, \autoref{sec:relatedwork} provides a literature review of the related work pertaining to Telegram, pandemics and disinformation dissemination. \autoref{sec:data} describes our technique for data collection from Telegram and establishes the terminology used in this paper. We segment the users into three groups -- the Disinformation Dozen, bot users, and human users -- and we describe the segmentation technique in \autoref{sec:useridentification}. We then describe our analytic methodology \autoref{sec:methdology}, discuss the results in \autoref{sec:discussion}, and finally state our conclusions \autoref{sec:conclusion}. 

\section{Related Work}
\label{sec:relatedwork}
\subsection{Telegram as a Mobile Messaging Platform}
Telegram is a cloud-based messaging application. There are applications for desktop computers as well as multiple mobile platforms, such as iPhone and Android. It is known for its end-to-end chat encryption. The messaging app was founded in 2013 by Russian brothers, and today is registered as a company in the British Virgin Islands and as an LLC in Dubai. As of 2023, Telegram has more than 700 million monthly active users worldwide \cite{forbesPavelDurov}.
Telegram is not only used as a messaging application but also as a platform for news media dissemination and consumption. For example, Telegram was found to be used by state-run media outlets to disseminate Persian language news to the Iranian public from 2015 to 2020 \cite{al2022news}. 

Telegram has been given close scrutiny by researchers and academic groups because it has been identified to be one of the most influential recruitment and planning tools used by terrorists and extremism groups \cite{walther2021us}. On top of that, Telegram is also actively used in politics. A study from 2020 reveals that Ukrainian politicians and their entourages maintained a series of Telegram channels that actively disseminated information relating to changing the political climate and discussing views on the Ukrainian government \cite{khaund2020telegram}. During the 2020 coronavirus pandemic, Telegram groups have been used to organize protests against control measures implemented by the German government \cite{weigand2022conspiracy} and mobilize white supremacists and hate \cite{guhl2020safe}.

Disinformation has been found to be present on Telegram, ranging from COVID-19 conspiracy theories to far-right disinformation topics \cite{walther2021us}, including QAnon conspiracy theories on the origins and of the coronavirus \cite{willaert2022disinformation}. Telegram channels impersonating celebrities or well-known services have been used to spread disinformation theories, with some messages reaching up to 1 million users \cite{la2021uncovering}. Unfortunately, these fake channels and the information they spread are difficult to identify even by the most media literate users, and therefore users are susceptible to believing these falsehoods as truth \cite{la2021uncovering}.

Telegram provides a rich source of information with its varied channels and discussions, and we harness this information to study human behavior with respect to the dissemination of coronavirus-related disinformation during the 2020 pandemic.

\subsection{Social Media and Pandemics}
During health pandemics, authorities use social media for diagnostic efforts and public communication \cite{liu2011organizations} or to access the public emotional state to aid in regional-level government decisions \cite{ng2020miss}. During the 2020 coronavirus pandemic, several governments set up their own social media channels to disseminate information and debunk fake news. Some governments leveraged Telegram channels for health-related communication. For example, the government of India used the channel ``MyGovCoronaNewsdesk" and the government of Singapore used the channel ``govsg".

While social media was used for crisis communication by authorities, mis- and disinformation on the coronavirus pandemic also spread quickly and wildly on these platforms. From over 10,000 falsehoods shared on several social media and mobile messaging platforms, it is clear that disinformation was a common global problem, but these narratives originated from a small number of individuals and organizations \cite{caliskan2023varieties}. There were also state-sponsored online disinformation campaigns that undermined socio-political systems, delegitimized public health and scientific bodies, and diverted public health responses, resulting in a combined cyber and biological pandemic \cite{bernard2021disinformation}.

Part of the disinformation spread was generated by uncertainty surrounding COVID-19 during the early stages of the pandemic, which led people to come up with cures from everyday foods \cite{ng2021coronavirus}. Conspiracy theories about the origins of the COVID-19 virus were also common. Due to the large volume of conspiracy theories, machine learning models have been constructed to systematically identify these theories at scale \cite{moffitt2021hunting}. 

Social media use during pandemics cuts both ways. The simplicity and the reach of the platforms enable governments and authorities to disseminate information relating to sanitary habits, governmental measures, and vaccination availability quickly and with little effort, using media that the public can easily access and in formats that are easily consumable \cite{zheng2013social}. On the other hand, the rich- and real-time information dissemination capability of social media allows malicious actors to thrive and disseminate disinformation at scale \cite{bernard2021disinformation}. The study of social media during pandemics is important in this digital age as a growing amount of information dissemination and consumption takes place online.

\subsection{Disinformation Dissemination}
During and after the COVID-19 pandemic, a string of studies identified patterns of dissemination and propagation of disinformation. Past work developed a Framework of Disinformation spread, characterizing how fake news could be disseminated through the official APIs of social media platforms and methodologically classifying the spread into four phases: network creation, profiling, content generation, and information dissemination \cite{ng2021does}. Disinformation has been found to transmit faster than real news, and some stories achieve sustainable propagation to achieve a substantially wider reach. In contrast, the volume of real news was found to drop drastically after it was initially posted \cite{pal2019propagation}.

To combat the propagation of disinformation, classification models have been constructed using recurrent and convolutional networks for early detection of disinformation in hopes of identifying and addressing these narratives early \cite{liu2018early}. These fake news detection approaches, including supervised machine learning techniques, multivariate time series models, and Bayesian learning models, aid in the detection of disinformation at scale\cite{hakak2020propagation}.

Bot accounts, or inauthentic and sometimes automated accounts, have also been observed to actively disseminate disinformation. Bot accounts exploit the online information ecosystem to sow misinformation by posting content and interacting with each other and human users through legitimate social connections \cite{shao2017spread}. A study of bot activity within the COVID-19 vaccine discourse identified that bots were ``hyper-social" users that were extremely active in contributing to low-credibility content distribution \cite{yuan2019examining}. Unfortunately, humans are largely unable to distinguish well designed social bots from genuine human accounts \cite{cresci2017paradigm}, because quantitatively, the features of bots and humans appear very similar \cite{ng2023botbuster}. Therefore, humans do unwittingly contribute to the spread of disinformation. For example, humans have been observed to retweet (i.e., share a post on X) the low-credibility content posted by bots and humans at the same rate \cite{shao2018spread}. 

Additionally, humans do not always react to falsehoods by debunking or investigating them. Past work shows that while humans express skepticism about disinformation posts \cite{ng2020analyzing}, they do not always take the time to investigate or debunk posts because they are either uninterested or believe it would take too long \cite{geeng2020fake}. Humans also share disinformation stories because they are sensational and exciting, thus collectively contributing to the widespread circulation of disinformation on a greater scale \cite{wen2014shut}.

Collectively, bots and human accounts contribute to widespread disinformation propagation. A disinformation campaign regarding the origin of COVID-19 as a synthesized virus from a biolab was discovered to be initiated by Russian state-funded groups and was spread through coordinated inauthentic amplification alongside the support of Russia's invasion of Ukraine among Russian-speaking users. At the same time, the narratives in this campaign were also naturally propagated by anti-vaccine and conservative English-speaking communities on X \cite{alieva2022investigating}.

Much of the past work in this space revolves around the spread of disinformation on the platform X, previously named Twitter, and our work builds on the concepts of past work to explore the spread of disinformation on Telegram.

\section{Data Collection and Processing}
\label{sec:data}
\begin{figure}[h!]
\centering
\includegraphics[width=0.7\textwidth]{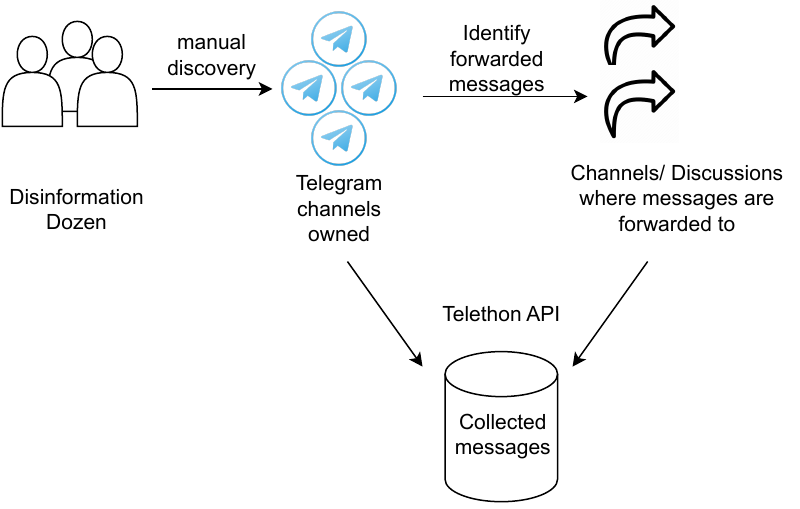}
\caption{Data Collection Pipeline}
\label{fig:data_collection_pipeline}
\end{figure}

In this section, we describe the data collection and processing pipelines. \autoref{fig:data_collection_pipeline} illustrates our data collection pipeline.

We collected messages posted to public Telegram channels via a snowball sampling method. We first referenced the names of the Disinformation Dozen released by the Center for Countering Digital Hate \cite{counterhateDisinformationDozen} and manually searched their names on Telegram search engines. Only eight of the Dozen had active accounts on Telegram at the time of this study. Channels are tools used to broadcast public messages to large audiences, but it is not possible to directly access lists of users who subscribe to these channels.  However, some channels also allow for comments or discussion which allows users to have public conversations about the channel content.  Of the 8 Disinformation Dozen accounts, 6 of them enabled discussions on their channels.  We collected both the channel content originating from the Disinformation Dozen and the associated discussion by other users.

We then computationally identified forwarded content in channel posts and discussions and identified which channels/users messages were forwarded from, forming our 1-hop channel list. We then collected channel messages from this 1-hop channel list. Using a 1-hop snowball sampling method provided a way of characterizing the dissemination of disinformation that stemmed from the original seed users -- the Disinformation Dozen -- as it identified users that disseminated messages through forwarding and collected information from the channels that received information second-hand. This technique allowed us to discover a large number of channels and users that were previously hidden from view.

To collect data from Telegram, we used the Python Telethon API\footnote{\url{https://docs.telethon.dev/en/stable/}}.  Using Python scripts, we scraped posts from the Disinformation Dozen Telegram channels, their associated groups, and any channels/groups that were linked to the initial set of channels/groups by a content forward. The resulting data is almost exclusively in the English language. We collected data from January to June 2023, obtaining a total of 7,711,975 messages from 10,633 channels that were written by 335,088 unique users.

In our analysis, we only accessed data from public channels and made no attempts to access private channels or chats. In terms of users, we only attempted to identify the accounts of the Disinformation Dozen. For the other users, we only observe public activity and we made no effort to identify these users beyond their public participation in the channels and time period stated above. In presenting our work, we redact user names so that the users are not able to be identified, further preserving user privacy.


\subsection{Terminology}
This study performs an examination across multiple Telegram channels. In \autoref{tab:terminology}, we define some of the terminology used in this study and the terminology unique to the social interactions on Telegram.

\begin{table}
    \centering
    \begin{tabular}{rl}
        \textbf{Terminology} & \textbf{Description} \\ \hline
        Disinformation Dozen & A set of twelve users identified to be the source of most coronavirus disinformation\\ 
        Bot users & Inauthentic users that are sometimes automated by computer software \\ 
        Human users & You and me \\ 
        Channel & A public broadcast list to disseminate messages to large audiences\\ 
        Discussion & A forum attached to a channel where users can post content\\
        Messages & A short text as a discrete unit of communication \\
        Forwarding & Sending a message to another channel using the ``forward" function \\
        Replying & Responding to a message within the same channel using the ``reply" function\\ 
        \hline
    \end{tabular}
    \caption{Terminology used in this paper}
    \label{tab:terminology}
\end{table}

\section{User Group Identification}
\label{sec:useridentification}
\begin{figure}[h!]
\includegraphics[width=1.0\textwidth]{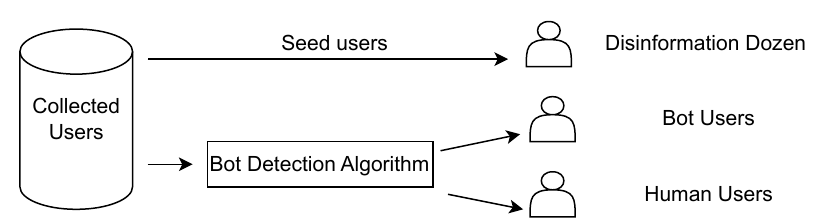}
\caption{User Group Identification Pipeline}
\label{fig:usergroupidentification}
\end{figure}

Within this study, we segment the users into three types: (1) the Disinformation Dozen identified by the Center for Countering Digital Hate as key spreaders of COVID disinformation \cite{counterhateDisinformationDozen}; (2) Bot Users as highly active inauthentic users; and (3) Human Users as authentic users of the platform. Within this paper, we study these three user groups, comparing the similarities and differences in terms of their message posting patterns, message topics, and network interactions. In this section, we describe each of the three types of users and how we methodologically identified these users. We begin by illustrating the user group identification pipeline in \autoref{fig:usergroupidentification}.

\subsection{Disinformation Dozen}
The CCDH published a report on March 24, 2021, establishing twelve users termed as the ``Disinformation Dozen", who are identified as being responsible for large-scale dissemination of COVID-19 disinformation and anti-vaccine claims circulating on social media platforms \cite{counterhateDisinformationDozen}.

This group of twelve people is mostly comprised of medical professionals and activists who have been found to create original disinformation content on X.  Their disinformation has been found to be more influential in conservative groups when compared to more liberal users \cite{nogara2022disinformation}. On Instagram, 96.13\% of their account pages redirect to the login page, suggesting that they might have been banned from the platform. Only 1.05\% of mentions of the account pages are replayable with complete post images \cite{bragg2023less}.

Studies surrounding the Disinformation Dozen have not been extensive, and this work contributes to current literature on the Disinformation Dozen by examining their role in the spread of information. We extend past work on the Disinformation Dozen by focusing on how disinformation propagated from their seed messages on Telegram rather than studying the content of their posts and/or the sources of information they relied upon \cite{nogara2022disinformation}.

We begin by describing the twelve influencers. 
\autoref{tab:disinfo_dozen} lists the names, occupations, and statuses on Telegram of the Disinformation Dozen. The names and occupations are adapted from \cite{nogara2022disinformation}. The profiles of these twelve people involve authoritative profiles such as physicians and alternative medicine activists. We note which of these dozen are ``active" on Telegram, indicating that we have discovered their profiles and channels. For those that we were unable to find, we list them as ``Not found" rather than ``Inactive" because it is possible they are operating under a different name on Telegram that we were unable to discover.

\begin{table}
    \centering
    \begin{tabular}{ccc}
        \textbf{Name} & \textbf{Occupation} & \textbf{Status} \\ \hline
        Joseph Mercola & osteopathic physician & Active \\ 
        Robert Kennedy Jr. & environment supporter & Not found \\ 
        Christiane Northrup & obstetrics and gynecology physician & Active \\ 
        Rashid Buttar & osteopathic physician & Active \\ 
        Erin Elizabeth & alternative medicine activist & Active \\ 
        Kelly Brogan & alternative medicine activist & Active \\ 
        Sayer Ji & alternative medicine activist & Active \\ 
        Kevin Jenkins & head of a no-vaccination group & Not found \\ 
        Sherri Tenpenny & osteopathic physician & Active \\ 
        Ben Tapper & chiropractor & Active \\ 
        Ty and Charlene Bollinger & alternative medicine activist & Not found \\ 
        Rizza Islam & anti-vaccination activist & Not found \\ 
        \hline
    \end{tabular}
    \caption{Name, occupation, and Telegram user status of the Disinformation Dozen in 2023}
    \label{tab:disinfo_dozen}
\end{table}

\subsection{Bot Users}
To detect bot users, we used the concept of bot detection, which defines ''bots" as user accounts that appear to be inauthentic. Bot detection algorithms typically make use of data from the user account, such as linguistic styles of user posts, user biography description, temporal information, and social network information to form feature sets used to differentiate bots from authentic accounts \cite{feng2022twibot}. These feature sets are fed into supervised machine learning algorithms, including Random Forests, Support Vector Machines, Logistic Regressions, Neural Networks, and Deep Learning methods \cite{heidari2021empirical}.

Bot users have been observed to spread and amplify disinformation \cite{shao2017spread}. For example, in the issue of the US Elections, bots on the social media platform X were observed spreading disinformation involving themes like ``Stop the Steal", manipulating public opinion with their online narratives \cite{chang2021social}. The 2020 coronavirus pandemic also revealed a series of bot campaigns that served to distort information \cite{himelein2021bots} and spread conspiracy theories \cite{moffitt2021hunting}.

Literature on Telegram bots generally focuses on techniques for developing bots for message exchange, chatting \cite{domashnev2019usage,de2016chatting} and monitoring server spaces \cite{idhom2018implementation}. An observational study analyzing the role and impact of Telegram bots in the Islamic State's online ecosystem revealed that Telegram bots identified through manual selection have two main roles: facilitating discussion and exchanging/augmenting content distribution efforts \cite{alrhmoun2023automating}.

We build on the idea that bots on Telegram can facilitate discussion and disinformation spread and identify the extent and impact of their activities on the stories put forth by the Disinformation Dozen.

\subsubsection{Bot User Identification through Bot Detection Algorithm}
In the absence of an existing robust bot detection methodology for Telegram, we adapted a bot detection algorithm that has been validated on both X (previously named Twitter) and Reddit.

To do so, we acquired a random subset of 3000 Telegram messages and a random subset of 3000 Tweets collected from X. The Tweets collected from X were collected during the same time period using the streaming API, filtered for the hashtags \#covid and \#vaccine. This dataset was previously used to analyze the changes in stance towards the coronavirus vaccine \cite{ng2022pro}.

With these datasets, we compared the messages, usernames and screen names, for these are the common fields between both platforms. \autoref{tab:diff_twitter_telegram} shows a statistics of comparison. We observed similarities in the average number of words and punctuation in messages, as well as the number of characters, numbers and capital letters in username and screen name. Further, we observed that there is no significant difference within the statistics at the $p<0.05$ level when we perform a two-tailed t-test between the statistics derived from X and Telegram. The length of the messages and the style of the usernames and screen names are similar between the two mediums. Therefore, we can use the bot detection algorithm tuned for the platform X for our Telegram data.

\begin{table}
    \centering
    \begin{tabular}{rcc}
         & \textbf{X} & \textbf{Telegram} \\ \hline 
         Number of words in messages & 10.09$\pm$32.96 & 18.35$\pm$33.39 \\
         Number of punctuation in messages & 5.26$\pm$8.34 & 3.98$\pm$8.31 \\
         Number of characters in username & 11.02$\pm$2.7 & 10.67$\pm$3.56 \\ 
         Numbers in username & 1.11$\pm$2.16 & 1.01$\pm$1.49 \\
         Capitals in username & 1.29$\pm$1.62 & 1.48$\pm$1.65 \\
         Numbers in screen name & 0.12$\pm$0.59 & 0.13$\pm$0.66 \\
         Capitals in screen name & 2.27$\pm$2.72 & 1.89$\pm$1.25 \\
         Number of words in screen name & 2.07$\pm$1.19 & 1.77$\pm$0.75 \\
         \hline 
    \end{tabular}
    \caption{Statistics of messages, username, and screenname between X and Telegram. All values between X and Telegram are statistically insignificant at the $p<0.05$ value by a two-tailed t-test, therefore the length of messages and the style of usernames and screen names are similar between the two mediums.}
    \label{tab:diff_twitter_telegram}
\end{table}

Because messages and user names on the Telegram messenger were similar to those on X, we chose to adapt a bot detection algorithm that has been trained on data from X to annotate our Telegram users called Botbuster \cite{ng2023botbuster}. The BotBuster algorithm uses a mixture-of-experts algorithm concept where each data field has its own prediction, and the predictions of all the data fields are aggregated together for the final bot prediction. The algorithm can take in a total of six input fields: user name, screen name, post, user metadata, and user description. This helps BotBuster handle the differing inputs from several social media platforms -- X, Instagram and Reddit. Since the Telegram platform does not have all the data fields that BotBuster can leverage, being able to activate a subset of experts to derive a final bot prediction is ideal for our Telegram dataset. BotBuster also facilitates reading input from a pre-collected dataset rather than pulling data live, allowing us to work on the data that we have collected. Finally, BotBuster does not identify bots based on a temporal point-of-view, therefore it reduces the problem of false positive detection on highly active human users. There may still be highly active human users that are detected as bots, but these users are likely to have features that are similar to bots.

We first harmonized the data field names extracted from Telegram to the BotBuster algorithm convention. Then, we ran the users that are not the Disinformation Dozen through the BotBuster algorithm. The algorithm provides a bot likelihood score between [0,1] that indicates the probability of the user being a bot. We use the threshold value of 0.5, where the user is determined to be a bot if the bot likelihood is equal to or above 0.5 and determined to be a human if the bot likelihood is below the 0.5 threshold value. With these parameters, the BotBuster algorithm determined our Telegram dataset to have 29.9\% bots and 70.1\% humans. 

\subsubsection{Annotation verification through Manual Annotation}
To verify the bot/human labels generated by the BotBuster algorithm, we performed manual annotation on a subset of data. From the BotBuster output, we randomly selected a 0.1\% sample by stratified sampling, thus ensuring that the proportion of bots/human users in the sample matches the proportions that are reflected by the BotBuster algorithm. In total, we extracted 2767 data points.

Two of the authors manually annotated the data points, reading through the user names and messages of each user before labeling the user as a bot or a human. In the event of a disagreement, a third annotator served to break the tie. All three annotators are native English speakers.

We also calculated the inter-annotator agreement between the first two annotators using Cohen's Kappa score. This score ranges from [-1, +1] and serves as an indication of the proportion of annotations that were not in agreement due to random chance. We only compared the first two annotators, with the third annotator serving as a tie-breaker. We obtained a Cohen Kappa score of 0.92. This score is close to 1, indicating sufficient agreement between the two annotators \cite{hallgren2012computing,artstein2008inter}. 

Finally, we harmonized the manual annotations through maximum pooling, taking the most common out of the three scores. These scores are thus regarded as the gold standard labels for this set of data. We compared the BotBuster-generated labels, finding a model F1-score of 72\%. This is a reasonable accuracy given that the original model that was fine-tuned on Twitter data performed with an F1 of 73\%.

From our observations, we determined a few characteristics of bot-like Telegram users. Note that while we present some examples in the list below, we present mild and neutral examples to avoid strong or disturbing language.
\begin{enumerate}
    \item Copying and pasting links in multiple messages, where the bulk of the links are the same except for the URL parameters and query strings; or keeps posting the exact same link
    \item Repeating the same message multiple times
    \item Writing in capital letters and the same number of exclamation marks for every message sent (e.g., ``THATS RIGHT, THANK YOU!!!")
    \item Repetitive and short messages (e.g., ``I agree with you", ``Inbox me please")
\end{enumerate}

\subsection{Human Users}
All other users that were not labeled manually as the Disinformation Dozen, nor algorithmically as a bot user, were labeled as Human Users.

\section{Methodology and Results}
\label{sec:methdology}
\begin{figure}[h!]
\includegraphics[width=1.0\textwidth]{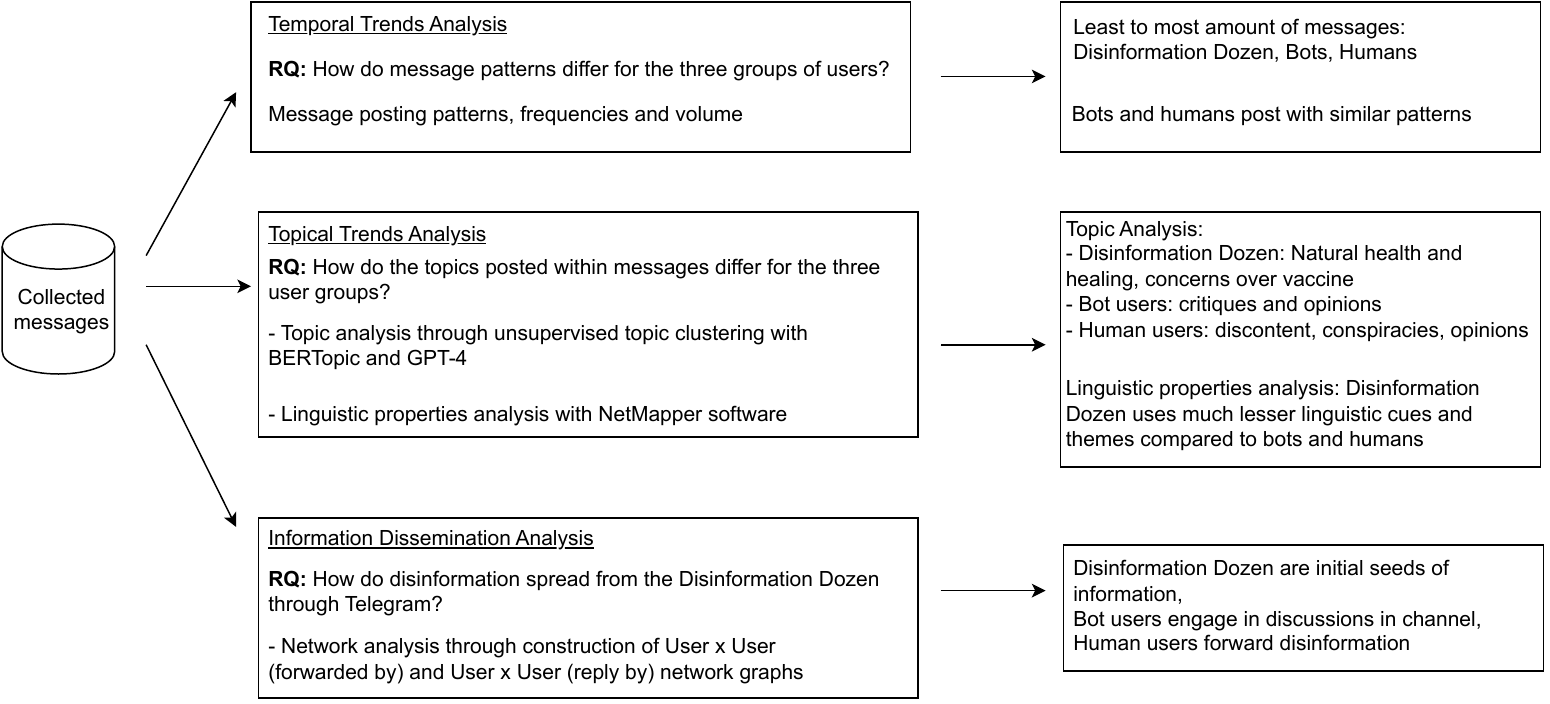}
\caption{Overview of Methodology and Brief Results}
\label{fig:methodology}
\end{figure}

With the Telegram data collected and annotated, we analyzed the data along three slices to answer the three research questions: temporal analysis, topical analysis, and network analysis. In this section, we detail our methodology and results for the three slices of analysis. We begin by illustrating our methodology and providing an overview of our results in \autoref{fig:methodology}.

\subsection{Temporal Trends}
\begin{figure}[h!]
\includegraphics[width=1.0\textwidth]{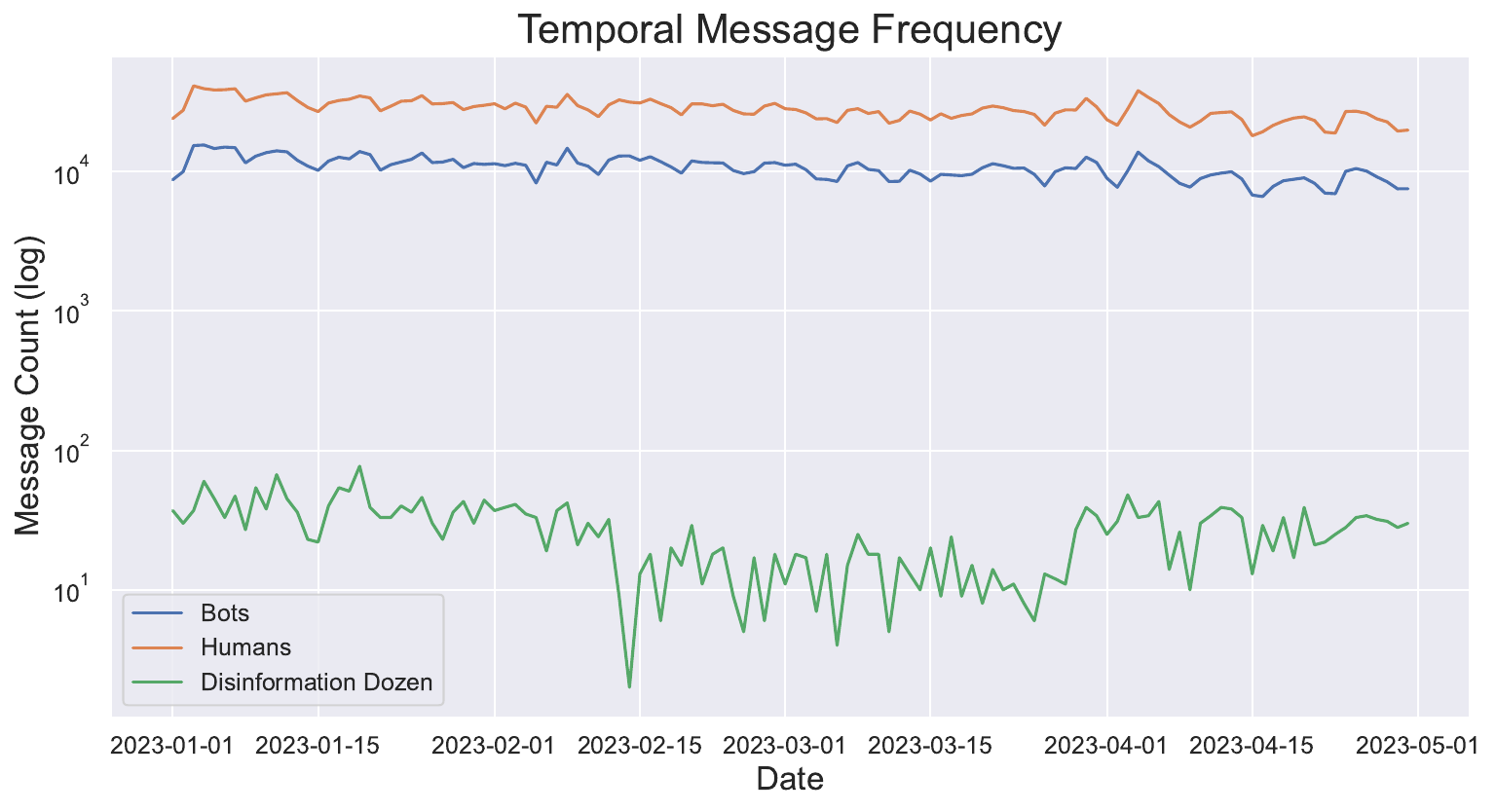}
\caption{Temporal Post Frequency per User Group. The Disinformation Dozen posts least frequently, while Humans post most frequently. Bots and humans post with similar frequency patterns, while the Disinformation Dozen posts the least.}
\label{fig:temporal}
\end{figure}

To answer Research Question 1 (How do message posting patterns differ for the three groups of users?), we use a temporal analysis approach. Temporal analysis provides an over-time understanding of the frequency of message sharing, showcasing the regularity and abnormalities of posting patterns across time. We plotted the frequency of posts of each of the three user groups across time, grouped by day, in \autoref{fig:temporal}.

From manual inspection, we observe that the posting patterns of Bots and Humans are extremely similar. We thus compare their similarity using a Pearson correlation metric\footnote{\url{https://docs.scipy.org/doc/scipy/reference/generated/scipy.stats.pearsonr.html}}. The Pearson correlation coefficient measures the linear relationship between the two posting frequency arrays as signals. It returns a value between [-1, +1]. Values closer to +1 indicate a positive correlation, values closer to -1 indicate a negative correlation, and values at 0 imply there is no correlation between the two signals. 

From the bot and human signals, we obtain a Pearson correlation value between the bot and human signals of 0.983, with a p-value of 4.05E-90. Since the p-value is less than 0.05, we conclude that the result is statistically significant and that the posting patterns of the bot and human users are positively correlated. That is, if there is an increase in the frequency of bot messages, there is also an increase in the frequency of human messages.


\subsection{Topical Trends}
To answer Research Question 2 (How do the topics posted within messages differ for the Disinformation Dozen, Bots, and Human users?), we use narrative analysis. We do so with two main techniques: topic analysis and linguistic analysis. 

For topic analysis, we developed topic clusters for each of the three user groups using unsupervised topic analysis. We used two popular preexisting algorithms in our methodology: BERTopic \cite{grootendorst2022bertopic} and OpenAI GPT-4 model \footnote{\url{https://platform.openai.com/docs/models/}} to extract topic sets from the messages. BERTopic returns a series of words that are most commonly used within a cluster of messages, while GPT provides a natural language interpretation of the messages as topics. For each user group, we input the set of messages that are authored by the users in the group into both topic clustering algorithms and extract the topics.

We derive the topic clusters for BERTopic by embedding each message into a vector using the all-MiniLM-L6-v2 transformer\footnote{\url{https://huggingface.co/sentence-transformers/all-MiniLM-L6-v2}} from HuggingFace sentence-transformer. These embeddings are used as input for a BERTopic model that uses a TF-IDF transformer and a hdbscan model to segregate embeddings into clusters. The model finally returns the key phrases that represent each topic cluster. For both algorithms, we evaluated topic groups using the kMeans clustering algorithm, setting the hyperparameter K to 40, which was established by the elbow method using the KneeLocator function.

We derive the topic clusters for GPT-4 in the following fashion. We first embed each message into a vector using the same all-MiniLM-L6-v2 sentence transformer. The vector embeddings are sorted into 40 clusters through the kMeans clustering algorithm.  We then extracted the text documents of each cluster and prompted a GPT-4 model with these documents. The prompt used is as follows: 

\begin{displayquote}
`````` \newline  I have a topic that contains the following documents: \{DOCUMENTS HERE\} \newline Based on the information above, extract a short (5 words or fewer) topic label in the following format: \newline topic: $<$topic label$>$ \newline
"""

\end{displayquote}

We present the top 5 topics by BERTopic and GPT labels on the three user group types in \autoref{tab:topics_disinfo_dozen}, \autoref{tab:topics_bots} and \autoref{tab:topics_humans}. We chose 5 topics as they represent the most salient topics determined through manual inspection. Both models generally output similar topic clusters, but the GPT labels present a more natural language interpretation. In general, the Disinformation Dozen posts messages related to vaccination, discouraging vaccination through medical discussions, vaccine safety and efficacy concerns, and unexpected deaths. Instead, they promote natural health and healing solutions. Bot users generally authored opinion messages, such as critiques towards the government, political affairs, and the coronavirus vaccine, suggesting that these bots were programmed to put forth messages consistent with specific ideologies. Human users focus on the Trump/Biden support, as well as other societal divides, such as racial and political divides. 

\begin{table}
    \centering
    \begin{tabular}{ccc}
       \textbf{S/N}  & \textbf{BERTopic Label} & \textbf{GPT label} \\ \hline 
       1 & greenmed health natural & Natural Health and Healing \\ 
       2 & dies suddenly & Unexpected Deaths and Causses \\ 
       3 & ios medium medical & Online Medical Discussion \\ 
       4 & covid masks dod & Covid masks and legal actions \\ 
       5 & covid vaccine vaers & Vaccine Safety and Efficacy Concerns \\ 
       \hline 
    \end{tabular}
    \caption{Generated topics for Disinformation Dozen}
    \label{tab:topics_disinfo_dozen}
\end{table}

\begin{table}
    \centering
    \begin{tabular}{ccc}
       \textbf{S/N}  & \textbf{BERTopic Label} & \textbf{GPT label} \\ \hline 
       1 & military justice law government & Government and Justice Critiques \\
       2 & biden trump president desantis & Political Affairs and opinions \\ 
       3 & channel telegram join & Social Media Conspiracy Discussions \\ 
       4 & vaccine jab covid & Covid Vaccine Opinions \\ 
       5 & mask wearing & Wearing Masks and Attitudes \\ 
       \hline 
    \end{tabular}
    \caption{Generated topics for Bot Users}
    \label{tab:topics_bots}
\end{table}

\begin{table}
    \centering
    \begin{tabular}{ccc}
       \textbf{S/N}  & \textbf{BERTopic Label} & \textbf{GPT label} \\ \hline 
       1 & trump biden president tucker & Trump Supporters, Against Biden \\ 
       2 & military white police people & Racial and Political Discontent \\ 
       3 & vaccine covid jabs & Covid and Vaccine Controversies \\ 
       4 & video channel rumble https & Conspiracy video links \\ 
       5 & mask wearing looks dog & Opinions on Masks and Appearances \\ 
       \hline 
    \end{tabular}
    \caption{Generated topics for Human Users}
    \label{tab:topics_humans}
\end{table}

We also performed a linguistic properties analysis using the Netmapper software\footnote{\url{http://www.casos.cs.cmu.edu/tools/data.php}}. The NetMapper software reads a message and counts the frequency of terms belonging to lexical categories, such as abusive, absolutist, positive, and negative terms, as well as terms relating to themes such as family, crime, and finance. These categories are built on a psycholinguistic theory that associates particular words with behavioral and cognitive states \cite{tausczik2010psychological}. We derive quantitative cues from NetMapper that represent linguistic properties and themes. We then calculate the mean and standard deviation of each of the cues per user group (Disinformation Dozen, bots, and humans) and compare the differences between the groups. 

\begin{figure}[h!]
\includegraphics[width=\textwidth]{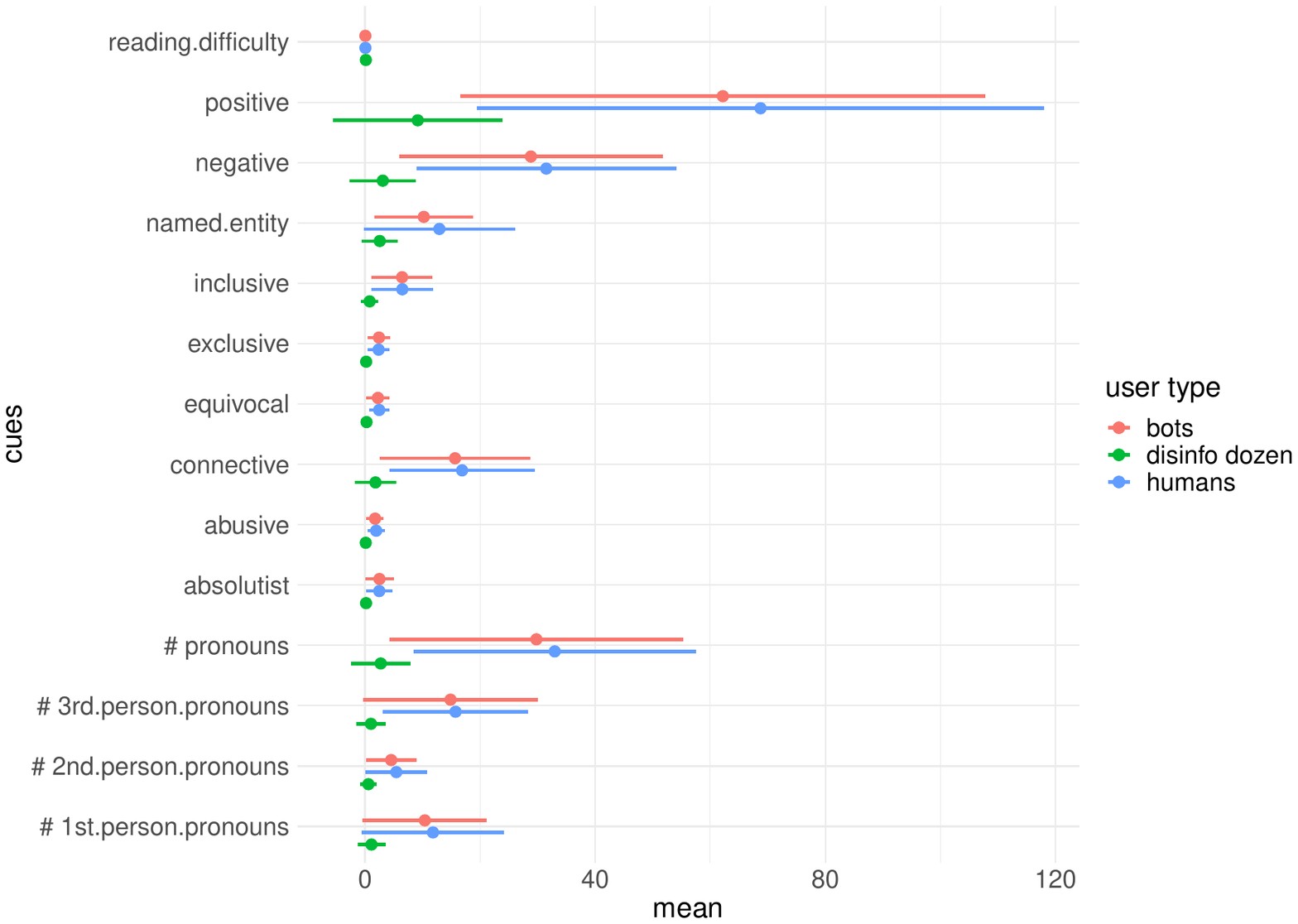}
\caption{Comparison of linguistic cues among the three user group}
\label{fig:netmapper_linguistic}
\end{figure}

\autoref{fig:netmapper_linguistic} shows the comparison of linguistic cues per user group. We find that the Flesch-Kincaid reading difficulty of the messages by all three groups are very low, for short online messages are not generally very complex. The largest use of linguistic cues is the positive and negative terms, which could be used in discouraging vaccination and encouraging natural health cures. Another commonly used linguistic feature is the 3rd person pronoun, e.g., ``we", which gives a sense of community and that we are all in it together \cite{ng2022pro}. The 1st person pronoun, e.g., ``I", is also frequently used, which provides a personal touch with personal anecdotes and opinions \cite{kacewicz2014pronoun}.

\begin{figure}[h!]
\includegraphics[width=1.0\textwidth]{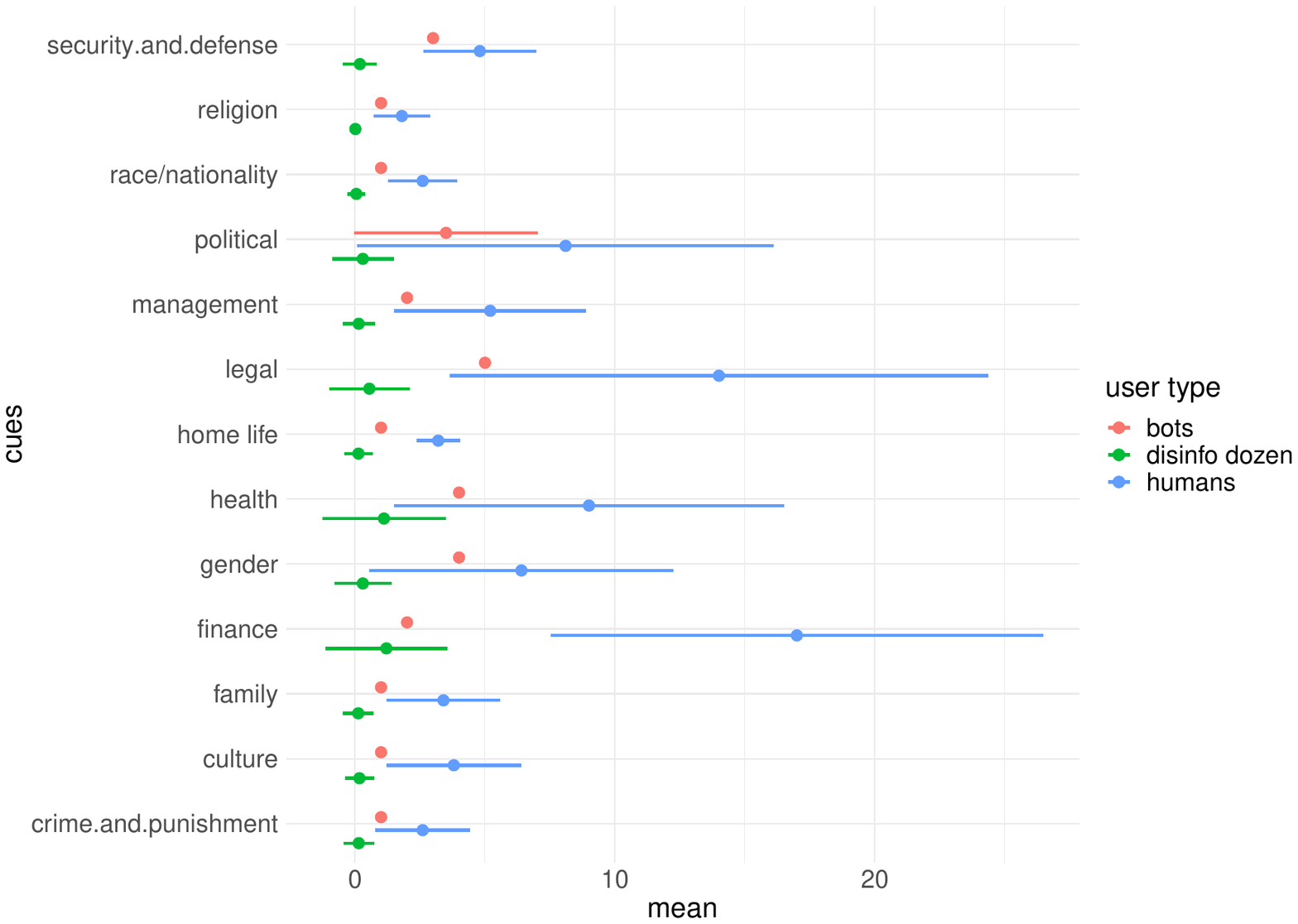}
\caption{Comparison of themes among the three user group}
\label{fig:netmapper_themes}
\end{figure}

\autoref{fig:netmapper_themes} shows the comparison of themes per user group. Human users have the most varied uses of themes, in particular, finance, health, political, and legal themes. This is followed by bot users, which tend to have a lot of political and legal themes, followed by health and gender themes. The Disinformation Dozen has the smallest use of theme-based words within their messages, but the most talked about theme is the health theme. This is likely because this group concentrates their messages around a focused theme and does not take part in side conversations.

\subsection{Information Dissemination Patterns}
We analyzed how information is disseminated in Telegram using a network analysis approach. This answers Research Question 3: How does disinformation begin spreading from the Disinformation Dozen throughout the Telegram messenger network? Network analysis provides us a bird's-eye view of the interactions between users and channels on Telegram \cite{willaert2022disinformation}. Visualizing user-to-user communication and interactions through a network perspective provides information about where the hubs and spokes of information flow are and the control of aggregation and dissemination of information.

We constructed two network diagrams: (a) User x User network by message forwarding and (b) User x User network by message reply. These two diagrams represent Telegram users as nodes and construct a user-to-user diagram (i.e., User x User). Two users are joined together by a link if they have an interaction between them. For the first diagram, the interaction is message forwarding; for the second diagram, the interaction is message replies. The width of the links between two users represents the frequency of the interaction that has occurred during the timeframe that we collected data. We also segregated bots and human users by means of a red and blue color scheme, respectively.

We used the betweenness centrality measure to size the nodes, where a larger node size corresponds with a larger betweenness centrality value. The betweenness centrality value indicates the extent to which a node lies on the path of information flow. It is normally calculated as a fraction of the shortest paths between node pairs that pass through a single node \cite{newman2005measure}. The higher the betweenness centrality of the node, the more information flows through it, and thus that user is influential as a hub for disseminating disinformation. Past work that studied the difference in betweenness centrality between bots and humans observed that bots are more likely to be on the bridges of information flow, reflecting higher betweenness centrality values \cite{ng2023combined}. Bots have been observed to occupy 7-16\% of the top users with the highest betweenness centrality ranking, implying that they play the role of bridging and mediating information diffusion \cite{cai2023network}. 

The network graphs and the betweenness centrality measures were calculated and plotted with the ORA software\footnote{\url{http://www.casos.cs.cmu.edu/projects/ora/software.php}}. We observe that the network constructed by messaging forwarding interactions, while sparse (network density = 5E-7), is dominated by human users, suggesting that human users are the user group that actively forward messages. The network constructed by messaging replying interactions is a lot denser (network density = 5.8E-5), which suggests heavier usage of the replying function compared to the forwarding function. That network is also dominated by bot users as influential users in the network, suggesting that bot users are highly active and generate lots of engagement for the coronavirus discourse.

\begin{figure}[h!]
\includegraphics[width=1.0\textwidth]{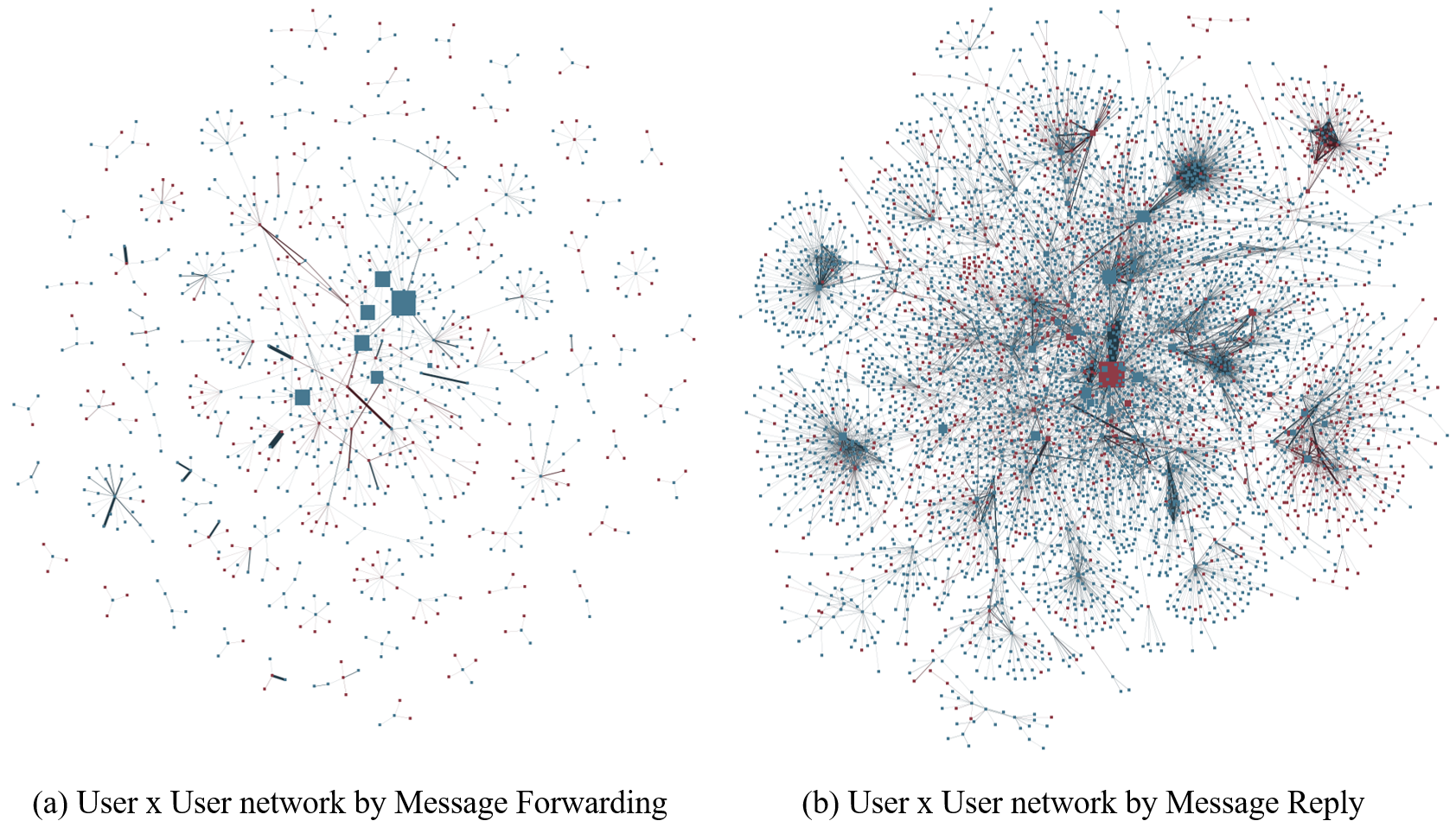}
\caption{Network graphs of User-User interaction with the Forward and Reply mechanism. The network is pruned to contain interaction links between users that have at least a frequency of 5. Blue nodes represent Human users, while Red nodes represent Bot users. The nodes are sized by the Betweenness Centrality measure. The Disinformation Dozen users are not observed in these graphs as they do not frequently forward other users' messages or reply to other users.}
\label{fig:network}
\end{figure}

For each of the two network graphs plotted, we calculated the average betweenness centrality values of the bot and human users and presented them in \autoref{tab:betweenness_centrality}. The quantitative calculations of the network centrality metric match the visual inspection of the network graphs: human users have higher betweenness centrality in the message-forwarding network, while bot users have higher betweenness centrality in the message-replying network. This suggests that in terms of message forwarding, human users are more likely to facilitate the flow of information, while in terms of message replies, bot users are more likely to accelerate discussions.

\begin{table}
    \centering
    \begin{tabular}{ccc}
        \textbf{User x User Network} & \textbf{Bot Users} & \textbf{Human Users} \\ \hline
        By Message Forwarding & 1.6E-4 $\pm$ 2.1E-3 &1.9E-4 $\pm$ 3.9E-3 \\
        By Message Reply & 1.5E-8 $\pm$ 2.7E-8 & 2.5E-10 $\pm$ 3.6E-8\\ \hline
    \end{tabular}
    \caption{Comparison of average Betweenness Centrality values between bot and human users}
    \label{tab:betweenness_centrality}
\end{table}

\section{Discussion}
\label{sec:discussion}
In this work, we studied the information dissemination patterns that stemmed from the Disinformation Dozen, a group of twelve users that the Center for Countering Digital Hate (CCDH) identified were key spreaders of COVID-19 disinformation online. We examined the disinformation spread through three lenses: temporal, topical, and network.

Our data collection timeline overlaps with the timeline in which the CCDH had performed their data collection and investigation. We posit that the CCDH performed their investigation during that time frame because it is likely the time frame that the Disinformation Dozen were highly active. 

We began our study by annotating bot users on Telegram, adapting a current bot detection algorithm to take in Telegram data as input. Given that the message length and user names are similar across the social media platform X and the Telegram messenger, we were able to adapt the current algorithm for use within this study. The similarity of messages shows that users, in general, posted with similar writing styles on both platforms. This finding provided justification for the use of tools that were developed for other social media platforms like X, Facebook, and Reddit, adapting them to an understudied and emerging platform. 

From the temporal lens, we observed that the Disinformation Dozen posted the fewest messages, followed by bots, than humans. The Disinformation Dozen were not the spreaders of disinformation, nor did they generate discourse. Rather, they were the originators of disinformation stories, which were propagated through the Telegram medium by bots and human users. 

From the topical lens, we found that the Disinformation Dozen posted messages on natural health and healing and concerns over the vaccine, bot users typically critique and post opinions, while human users posted their opinions, discontent and conspiracy theories. From these topical divides, we posit that Disinformation Dozen was trying to promote natural health cures over the coronavirus vaccination. With this specific information range, the topics put forth by the Disinformation Dozen were the narrowest, and they used the least varied linguistic cues. Bot users were programmed to post critical remarks and opinions. Due to their range of topics, from political affairs to governments to social media, they had a variety of themes and linguistic styles within their messages. Human users talked generally about their opinions and discontent, therefore they touched on the widest range of topics and used the most varied linguistic styles. 

From the network analysis lens, we observed that the Disinformation Dozen were not actually the main users involved in the propagation of disinformation on Telegram but rather drove the initial information dissemination. They were able to command large followings with their channels, mainly because of their background as established medical professionals or activists. This demonstrates a real-world example of authority bias, a type of behavioral bias where the opinions and instructions of authority figures are unquestioned \cite{howard2019bandwagon}. Authority bias is common in the medical field with authority figures, professional experts, and hierarchy. It can be seen in the doctor-patient relationship, which many of the Disinformation Dozen mirrors \cite{silvester2021authority}. Therefore, using their medical expertise as an authority, the Disinformation Dozen seeded their channels and allowed other users to generate discourse and forward information. 

These forwarding users included bot users who were active using the platform's replying mechanic, creating a huge amount of engagement around COVID-19 disinformation within the original channels. The user set also includes human users who were active in forwarding disinformation between platforms. Whether they forwarded this information because they truly believed the disinformation or because they found it amusing \cite{duffy2020too}, the act of forwarding propagates the disinformation to another set of users on other channels. We posit that the bots may have participated more in conversations within the channel rather than forwarding information because their software programming was set to analyze conversations from and reply to only one specific channel. 

The bot behavior we observe on Telegram is somewhat surprising in that it is distinct from common bot behavior on other platforms, like X.  In other social media, it is common to find bots functioning as bridges between users and groups (e.g., by using @ mentions to link users) \cite{gilani2017bots,samper2021bot}. In many cases, these bridging bots do not post content at all. This paradigm is completely reversed on Telegram, with bots serving primarily as content contributors and not regularly attempting to bridge users/groups.

Discovering content on Telegram is much more constrained than it is on other platforms such as X or Reddit because there is no algorithmic news feed or any content discovery mechanism beyond message forwarding.  Interestingly, this constraint seems to impact bot behavior, and it suggests that bots may have more limited capability to influence social networks on Telegram than on platforms that have greater support for discovery and connection-making.

The information creation and sharing behavior of the Disinformation Dozen, bots, and humans were similarly observed in the platform X by past works. A previous study observed how the Disinformation Dozen acted as sources of information by creating original content while interacting very little with other users \cite{nogara2022disinformation}. Meanwhile, other users had a fair amount of user-to-user interaction and substantially forwarded content.

We also observed separated clusters of communities in the network diagrams, as well as clusters connected together by bridging users. This is similar to past work observing patterns of information dissemination on Telegram. While communities on Telegram tend to interact and communicate within their own groups, there is substantial information sharing between communities and ideologies through the forwarding mechanism, facilitating information spread \cite{kloo2023social}.

While the volume and virality of disinformation on online platforms are alarming, we take comfort in studies that show that the social media platforms are taking action to remove such posts: less than 4\% of the Instagram posts by the Disinformation Dozen can be replayed, as they have been removed from the archive by the Instagram platform \cite{bragg2023less}. However, it is much more difficult to police sets of conversational channels in the Telegram messenger setup, and our work shows that many of the messages are still alive up to six months after the post dates. While Telegram has shown to have removed and banned some posts containing disinformation and conspiracy theories, many cloned channels and messages appear quickly, suggesting further work is required to moderate the messenger \cite{la2021uncovering}.

\subsection{Limitations and Future Work}
Our work is not without its limitations, and thus one should exercise caution before using it to inform further policy. We present some of the limitations in this section and the avenues for future work.

Finding channels of the Disinformation Dozen is relatively easy because this group of users wants to be found. They want to broadcast their (dis)information to as large an audience as possible and thus use similar user names and descriptions across social media platforms. Still, we did not manage to find all the dozen on Telegram, and the presence of which might bring results that are different from the ones observed. We were only able to locate 8 of the 12 Disinformation Dozen on Telegram. It is possible that the other 4 individuals have Telegram accounts that are not identifiable but are still relevant in the medical disinformation space. We expect that our snowball sampling approach would have found these accounts (and it very well may have), but we do not have any way to verify the identities of these users. Additionally, the snowball sampling method might introduce bias in the data collection. The snowball is constructed by the conversation within the channels manned by the Disinformation Dozen and, therefore, reflects users who are actively involved with the twelve original users. These issues might limit the breadth of the study's insights, therefore caution should be used in extrapolating our results, as there may be users and channels that separately spread disinformation and propagate anti-vaccine ideologies but are not directly linked to our initial set of seed channels.

While our bot detection algorithm has achieved an accuracy of 72\% on the dataset, the algorithm used was trained on a series of datasets from the platform X. Tweets from X, while similar to Telegram messages, are not exactly the same, and therefore to further improve the accuracy, the algorithm could be modified to be trained directly on Telegram datasets, in order to better fine-tune the algorithm to handle Telegram-specific data. 

Finally, our study focuses on data primarily in the English language. Telegram is used by users all around the world, and therefore disinformation can be propagated in multiple languages. Future work would be to investigate the propagation of coronavirus disinformation that stems from the Disinformation Dozen in languages other than English. Such work will provide insight into the global reach of the disinformation propagation.

\section{Conclusion}
\label{sec:conclusion}
The fear of the unknown and the doubts about the efficacy of the vaccine, combined with government-imposed isolation generated by the COVID-19 pandemic, created a fertile environment for disinformation to flourish.

The spread of disinformation online is a concerning phenomenon, especially when it involves public health, as it weakens the government's abilities to control and eradicate the pandemic \cite{caliskan2023varieties}. In this study, we examined the spread of COVID-19 disinformation that stemmed from an influential group of disinformation spreaders on Telegram by identifying and studying the roles of three user group types: the Disinformation Dozen, bot users, and human users. 

Contrary to the report by CCDH, we found that the Disinformation Dozen were not the most prolific in spreading coronavirus-related and anti-vaccination disinformation on Telegram. Instead, the Disinformation Dozen initiated disinformation themes, while bot users were essential in sustaining the conversations, and human users were crucial in disseminating information to other Telegram channels.

This study also demonstrated some unique and unexpected human and bot behavior on Telegram compared to other social media.  Specifically, we found that community bridging behavior was primarily human-driven on Telegram. Additionally, we showed that a large community centered on medical disinformation was able to form on Telegram in spite of the platform's somewhat limited functionality for users to discover new content.  These differences highlight the importance of future study into Telegram (and other similar platforms). The research community has focused on platforms like X over the past 10 years due to ease of data access, but our findings suggest these studies may not generalize as well as previously thought.  Moreover, with platforms like X and Reddit no longer supporting the research community through platform-provided APIs, Telegram is becoming a more useful platform for future work.

Within this work, we developed a repeatable methodology for analyzing Telegram messages and users across temporal, topical, and network domains. We hope our work motivates and serves as a springboard for future analysis of disinformation propagation on Telegram and other less-studied social media platforms.

\section{Declarations}
\subsection{Funding}
This material is based upon work supported by the U.S. Army Research Office and the U.S. Army Futures Command under Contract No. W911NF-20-D-0002, Office of Naval Research (Bothunter, N000141812108), US Army Scalable Technologies for Social Cybersecurity (W911NF20D0002), Air Force Research Laboratory/CyberFit (FA86502126244), Office of Naval Research Scalable Tools for Social Media Assessment (N000142112229). The content of the information does not necessarily reflect the position or the policy of the government and no official endorsement should be inferred.

\subsection{IRB Declaration}

\subsection{Conflict of Interest/ Competing Interests}
The authors declare that they have no conflict of interests

\subsection{Availability of data and code}
Data and code can be obtained from contacting the corresponding author, in accordance to the sharing policy of Telegram.

\subsection{Authors' contributions}
Lynnette conceptualized the study and performed the analysis and writing. Ian collected the data and performed the analysis. Samantha performed data annotation and analysis. Kathleen reviewed the work. All authors have agreed upon the manuscript.

\bibliography{sn-bibliography}


\begin{thebibliography}{57}
\ifx \bisbn   \undefined \def \bisbn  #1{ISBN #1}\fi
\ifx \binits  \undefined \def \binits#1{#1}\fi
\ifx \bauthor  \undefined \def \bauthor#1{#1}\fi
\ifx \batitle  \undefined \def \batitle#1{#1}\fi
\ifx \bjtitle  \undefined \def \bjtitle#1{#1}\fi
\ifx \bvolume  \undefined \def \bvolume#1{\textbf{#1}}\fi
\ifx \byear  \undefined \def \byear#1{#1}\fi
\ifx \bissue  \undefined \def \bissue#1{#1}\fi
\ifx \bfpage  \undefined \def \bfpage#1{#1}\fi
\ifx \blpage  \undefined \def \blpage #1{#1}\fi
\ifx \burl  \undefined \def \burl#1{\textsf{#1}}\fi
\ifx \doiurl  \undefined \def \doiurl#1{\url{https://doi.org/#1}}\fi
\ifx \betal  \undefined \def \betal{\textit{et al.}}\fi
\ifx \binstitute  \undefined \def \binstitute#1{#1}\fi
\ifx \binstitutionaled  \undefined \def \binstitutionaled#1{#1}\fi
\ifx \bctitle  \undefined \def \bctitle#1{#1}\fi
\ifx \beditor  \undefined \def \beditor#1{#1}\fi
\ifx \bpublisher  \undefined \def \bpublisher#1{#1}\fi
\ifx \bbtitle  \undefined \def \bbtitle#1{#1}\fi
\ifx \bedition  \undefined \def \bedition#1{#1}\fi
\ifx \bseriesno  \undefined \def \bseriesno#1{#1}\fi
\ifx \blocation  \undefined \def \blocation#1{#1}\fi
\ifx \bsertitle  \undefined \def \bsertitle#1{#1}\fi
\ifx \bsnm \undefined \def \bsnm#1{#1}\fi
\ifx \bsuffix \undefined \def \bsuffix#1{#1}\fi
\ifx \bparticle \undefined \def \bparticle#1{#1}\fi
\ifx \barticle \undefined \def \barticle#1{#1}\fi
\bibcommenthead
\ifx \bconfdate \undefined \def \bconfdate #1{#1}\fi
\ifx \botherref \undefined \def \botherref #1{#1}\fi
\ifx \url \undefined \def \url#1{\textsf{#1}}\fi
\ifx \bchapter \undefined \def \bchapter#1{#1}\fi
\ifx \bbook \undefined \def \bbook#1{#1}\fi
\ifx \bcomment \undefined \def \bcomment#1{#1}\fi
\ifx \oauthor \undefined \def \oauthor#1{#1}\fi
\ifx \citeauthoryear \undefined \def \citeauthoryear#1{#1}\fi
\ifx \endbibitem  \undefined \def \endbibitem {}\fi
\ifx \bconflocation  \undefined \def \bconflocation#1{#1}\fi
\ifx \arxivurl  \undefined \def \arxivurl#1{\textsf{#1}}\fi
\csname PreBibitemsHook\endcsname

\bibitem[\protect\citeauthoryear{Ng and Carley}{2021}]{ng2021coronavirus}
\begin{barticle}
\bauthor{\bsnm{Ng}, \binits{L.H.X.}},
\bauthor{\bsnm{Carley}, \binits{K.M.}}:
\batitle{“the coronavirus is a bioweapon”: classifying coronavirus stories on fact-checking sites}.
\bjtitle{Computational and Mathematical Organization Theory}
\bvolume{27}(\bissue{2}),
\bfpage{179}--\blpage{194}
(\byear{2021})
\end{barticle}
\endbibitem

\bibitem[\protect\citeauthoryear{Uscinski et~al.}{2020}]{uscinski2020people}
\begin{botherref}
\oauthor{\bsnm{Uscinski}, \binits{J.E.}},
\oauthor{\bsnm{Enders}, \binits{A.M.}},
\oauthor{\bsnm{Klofstad}, \binits{C.}},
\oauthor{\bsnm{Seelig}, \binits{M.}},
\oauthor{\bsnm{Funchion}, \binits{J.}},
\oauthor{\bsnm{Everett}, \binits{C.}},
\oauthor{\bsnm{Wuchty}, \binits{S.}},
\oauthor{\bsnm{Premaratne}, \binits{K.}},
\oauthor{\bsnm{Murthi}, \binits{M.}}:
Why do people believe covid-19 conspiracy theories?
Harvard Kennedy School Misinformation Review
\textbf{1}(3)
(2020)
\end{botherref}
\endbibitem

\bibitem[\protect\citeauthoryear{CCDH}{2021}]{counterhateDisinformationDozen}
\begin{botherref}
\oauthor{\bsnm{CCDH}}:
{T}he {D}isinformation {D}ozen — {C}enter for {C}ountering {D}igital {H}ate | {C}{C}{D}{H} --- counterhate.com.
\url{https://counterhate.com/research/the-disinformation-dozen/}.
[Accessed 25-10-2023]
(2021)
\end{botherref}
\endbibitem

\bibitem[\protect\citeauthoryear{Nogara et~al.}{2022}]{nogara2022disinformation}
\begin{bchapter}
\bauthor{\bsnm{Nogara}, \binits{G.}},
\bauthor{\bsnm{Vishnuprasad}, \binits{P.S.}},
\bauthor{\bsnm{Cardoso}, \binits{F.}},
\bauthor{\bsnm{Ayoub}, \binits{O.}},
\bauthor{\bsnm{Giordano}, \binits{S.}},
\bauthor{\bsnm{Luceri}, \binits{L.}}:
\bctitle{The disinformation dozen: An exploratory analysis of covid-19 disinformation proliferation on twitter}.
In: \bbtitle{Proceedings of the 14th ACM Web Science Conference 2022},
pp. \bfpage{348}--\blpage{358}
(\byear{2022})
\end{bchapter}
\endbibitem

\bibitem[\protect\citeauthoryear{Krishnan et~al.}{2021}]{krishnan2021research}
\begin{barticle}
\bauthor{\bsnm{Krishnan}, \binits{N.}},
\bauthor{\bsnm{Gu}, \binits{J.}},
\bauthor{\bsnm{Tromble}, \binits{R.}},
\bauthor{\bsnm{Abroms}, \binits{L.C.}}:
\batitle{Research note: Examining how various social media platforms have responded to covid-19 misinformation}.
\bjtitle{Harvard Kennedy School Misinformation Review}
\bvolume{2}(\bissue{6}),
\bfpage{1}--\blpage{25}
(\byear{2021})
\end{barticle}
\endbibitem

\bibitem[\protect\citeauthoryear{Forbes}{2023}]{forbesPavelDurov}
\begin{botherref}
\oauthor{\bsnm{Forbes}}:
{P}avel {D}urov --- forbes.com.
\url{https://www.forbes.com/profile/pavel-durov/?sh=77a6811e14c5}.
[Accessed 26-10-2023]
(2023)
\end{botherref}
\endbibitem

\bibitem[\protect\citeauthoryear{Ng and Loke}{2020}]{ng2020analyzing}
\begin{barticle}
\bauthor{\bsnm{Ng}, \binits{L.H.X.}},
\bauthor{\bsnm{Loke}, \binits{J.Y.}}:
\batitle{Analyzing public opinion and misinformation in a covid-19 telegram group chat}.
\bjtitle{IEEE Internet Computing}
\bvolume{25}(\bissue{2}),
\bfpage{84}--\blpage{91}
(\byear{2020})
\end{barticle}
\endbibitem

\bibitem[\protect\citeauthoryear{Willaert et~al.}{2022}]{willaert2022disinformation}
\begin{botherref}
\oauthor{\bsnm{Willaert}, \binits{T.}},
\oauthor{\bsnm{Peeters}, \binits{S.}},
\oauthor{\bsnm{Seijbel}, \binits{J.}},
\oauthor{\bsnm{Van~Raemdonck}, \binits{N.}}:
Disinformation networks: a quali-quantitative investigation of antagonistic dutch-speaking telegram channels.
First Monday
(2022)
\end{botherref}
\endbibitem

\bibitem[\protect\citeauthoryear{Sosa and Sharoff}{2022}]{sosa2022multimodal}
\begin{bchapter}
\bauthor{\bsnm{Sosa}, \binits{J.}},
\bauthor{\bsnm{Sharoff}, \binits{S.}}:
\bctitle{Multimodal pipeline for collection of misinformation data from telegram}.
In: \bbtitle{Proceedings of the Thirteenth Language Resources and Evaluation Conference},
pp. \bfpage{1480}--\blpage{1489}
(\byear{2022})
\end{bchapter}
\endbibitem

\bibitem[\protect\citeauthoryear{Al-Rawi}{2022}]{al2022news}
\begin{barticle}
\bauthor{\bsnm{Al-Rawi}, \binits{A.}}:
\batitle{News loopholing: Telegram news as portable alternative media}.
\bjtitle{Journal of Computational Social Science}
\bvolume{5}(\bissue{1}),
\bfpage{949}--\blpage{968}
(\byear{2022})
\end{barticle}
\endbibitem

\bibitem[\protect\citeauthoryear{Walther and McCoy}{2021}]{walther2021us}
\begin{barticle}
\bauthor{\bsnm{Walther}, \binits{S.}},
\bauthor{\bsnm{McCoy}, \binits{A.}}:
\batitle{Us extremism on telegram}.
\bjtitle{Perspectives on Terrorism}
\bvolume{15}(\bissue{2}),
\bfpage{100}--\blpage{124}
(\byear{2021})
\end{barticle}
\endbibitem

\bibitem[\protect\citeauthoryear{Khaund et~al.}{2020}]{khaund2020telegram}
\begin{bchapter}
\bauthor{\bsnm{Khaund}, \binits{T.}},
\bauthor{\bsnm{Hussain}, \binits{M.N.}},
\bauthor{\bsnm{Shaik}, \binits{M.}},
\bauthor{\bsnm{Agarwal}, \binits{N.}}:
\bctitle{Telegram: Data collection, opportunities and challenges}.
In: \bbtitle{Annual International Conference on Information Management and Big Data},
pp. \bfpage{513}--\blpage{526}
(\byear{2020}).
\bcomment{Springer}
\end{bchapter}
\endbibitem

\bibitem[\protect\citeauthoryear{Weigand et~al.}{2022}]{weigand2022conspiracy}
\begin{bchapter}
\bauthor{\bsnm{Weigand}, \binits{M.}},
\bauthor{\bsnm{Weber}, \binits{M.}},
\bauthor{\bsnm{Gruber}, \binits{J.}}:
\bctitle{Conspiracy narratives in the protest movement against covid-19 restrictions in germany. a long-term content analysis of telegram chat groups.}
In: \bbtitle{Proceedings of the Fifth Workshop on Natural Language Processing and Computational Social Science (NLP+ CSS)},
pp. \bfpage{52}--\blpage{58}
(\byear{2022})
\end{bchapter}
\endbibitem

\bibitem[\protect\citeauthoryear{Guhl and Davey}{2020}]{guhl2020safe}
\begin{botherref}
\oauthor{\bsnm{Guhl}, \binits{J.}},
\oauthor{\bsnm{Davey}, \binits{J.}}:
A safe space to hate: White supremacist mobilisation on telegram.
Institute for Strategic Dialogue
\textbf{26}
(2020)
\end{botherref}
\endbibitem

\bibitem[\protect\citeauthoryear{La~Morgia et~al.}{2021}]{la2021uncovering}
\begin{botherref}
\oauthor{\bsnm{La~Morgia}, \binits{M.}},
\oauthor{\bsnm{Mei}, \binits{A.}},
\oauthor{\bsnm{Mongardini}, \binits{A.M.}},
\oauthor{\bsnm{Wu}, \binits{J.}}:
Uncovering the dark side of telegram: Fakes, clones, scams, and conspiracy movements.
arXiv preprint arXiv:2111.13530
(2021)
\end{botherref}
\endbibitem

\bibitem[\protect\citeauthoryear{Liu and Kim}{2011}]{liu2011organizations}
\begin{barticle}
\bauthor{\bsnm{Liu}, \binits{B.F.}},
\bauthor{\bsnm{Kim}, \binits{S.}}:
\batitle{How organizations framed the 2009 h1n1 pandemic via social and traditional media: Implications for us health communicators}.
\bjtitle{Public relations review}
\bvolume{37}(\bissue{3}),
\bfpage{233}--\blpage{244}
(\byear{2011})
\end{barticle}
\endbibitem

\bibitem[\protect\citeauthoryear{Ng et~al.}{2020}]{ng2020miss}
\begin{bchapter}
\bauthor{\bsnm{Ng}, \binits{H.X.L.}},
\bauthor{\bsnm{Lee}, \binits{R.K.-W.}},
\bauthor{\bsnm{Awal}, \binits{M.R.}}:
\bctitle{I miss you babe: Analyzing emotion dynamics during covid-19 pandemic}.
In: \bbtitle{Proceedings of the Fourth Workshop on Natural Language Processing and Computational Social Science},
pp. \bfpage{41}--\blpage{49}
(\byear{2020})
\end{bchapter}
\endbibitem

\bibitem[\protect\citeauthoryear{Caliskan and Kilicaslan}{2023}]{caliskan2023varieties}
\begin{barticle}
\bauthor{\bsnm{Caliskan}, \binits{C.}},
\bauthor{\bsnm{Kilicaslan}, \binits{A.}}:
\batitle{Varieties of corona news: a cross-national study on the foundations of online misinformation production during the covid-19 pandemic}.
\bjtitle{Journal of Computational Social Science}
\bvolume{6}(\bissue{1}),
\bfpage{191}--\blpage{243}
(\byear{2023})
\end{barticle}
\endbibitem

\bibitem[\protect\citeauthoryear{Bernard et~al.}{2021}]{bernard2021disinformation}
\begin{barticle}
\bauthor{\bsnm{Bernard}, \binits{R.}},
\bauthor{\bsnm{Bowsher}, \binits{G.}},
\bauthor{\bsnm{Sullivan}, \binits{R.}},
\bauthor{\bsnm{Gibson-Fall}, \binits{F.}}:
\batitle{Disinformation and epidemics: anticipating the next phase of biowarfare}.
\bjtitle{Health security}
\bvolume{19}(\bissue{1}),
\bfpage{3}--\blpage{12}
(\byear{2021})
\end{barticle}
\endbibitem

\bibitem[\protect\citeauthoryear{Moffitt et~al.}{2021}]{moffitt2021hunting}
\begin{barticle}
\bauthor{\bsnm{Moffitt}, \binits{J.}},
\bauthor{\bsnm{King}, \binits{C.}},
\bauthor{\bsnm{Carley}, \binits{K.M.}}:
\batitle{Hunting conspiracy theories during the covid-19 pandemic}.
\bjtitle{Social Media+ Society}
\bvolume{7}(\bissue{3}),
\bfpage{20563051211043212}
(\byear{2021})
\end{barticle}
\endbibitem

\bibitem[\protect\citeauthoryear{Zheng}{2013}]{zheng2013social}
\begin{barticle}
\bauthor{\bsnm{Zheng}, \binits{L.}}:
\batitle{Social media in chinese government: Drivers, challenges and capabilities}.
\bjtitle{Government information quarterly}
\bvolume{30}(\bissue{4}),
\bfpage{369}--\blpage{376}
(\byear{2013})
\end{barticle}
\endbibitem

\bibitem[\protect\citeauthoryear{Ng and Taeihagh}{2021}]{ng2021does}
\begin{barticle}
\bauthor{\bsnm{Ng}, \binits{L.H.}},
\bauthor{\bsnm{Taeihagh}, \binits{A.}}:
\batitle{How does fake news spread? understanding pathways of disinformation spread through apis}.
\bjtitle{Policy \& Internet}
\bvolume{13}(\bissue{4}),
\bfpage{560}--\blpage{585}
(\byear{2021})
\end{barticle}
\endbibitem

\bibitem[\protect\citeauthoryear{Pal and Chua}{2019}]{pal2019propagation}
\begin{bchapter}
\bauthor{\bsnm{Pal}, \binits{A.}},
\bauthor{\bsnm{Chua}, \binits{A.Y.}}:
\bctitle{Propagation pattern as a telltale sign of fake news on social media}.
In: \bbtitle{2019 5th International Conference on Information Management (ICIM)},
pp. \bfpage{269}--\blpage{273}
(\byear{2019}).
\bcomment{IEEE}
\end{bchapter}
\endbibitem

\bibitem[\protect\citeauthoryear{Liu and Wu}{2018}]{liu2018early}
\begin{bchapter}
\bauthor{\bsnm{Liu}, \binits{Y.}},
\bauthor{\bsnm{Wu}, \binits{Y.-F.}}:
\bctitle{Early detection of fake news on social media through propagation path classification with recurrent and convolutional networks}.
In: \bbtitle{Proceedings of the AAAI Conference on Artificial Intelligence},
vol. \bseriesno{32}
(\byear{2018})
\end{bchapter}
\endbibitem

\bibitem[\protect\citeauthoryear{Hakak et~al.}{2020}]{hakak2020propagation}
\begin{bchapter}
\bauthor{\bsnm{Hakak}, \binits{S.}},
\bauthor{\bsnm{Khan}, \binits{W.Z.}},
\bauthor{\bsnm{Bhattacharya}, \binits{S.}},
\bauthor{\bsnm{Reddy}, \binits{G.T.}},
\bauthor{\bsnm{Choo}, \binits{K.-K.R.}}:
\bctitle{Propagation of fake news on social media: challenges and opportunities}.
In: \bbtitle{Computational Data and Social Networks: 9th International Conference, CSoNet 2020, Dallas, TX, USA, December 11--13, 2020, Proceedings 9},
pp. \bfpage{345}--\blpage{353}
(\byear{2020}).
\bcomment{Springer}
\end{bchapter}
\endbibitem

\bibitem[\protect\citeauthoryear{Shao et~al.}{2017}]{shao2017spread}
\begin{barticle}
\bauthor{\bsnm{Shao}, \binits{C.}},
\bauthor{\bsnm{Ciampaglia}, \binits{G.L.}},
\bauthor{\bsnm{Varol}, \binits{O.}},
\bauthor{\bsnm{Flammini}, \binits{A.}},
\bauthor{\bsnm{Menczer}, \binits{F.}}:
\batitle{The spread of fake news by social bots}.
\bjtitle{arXiv preprint arXiv:1707.07592}
\bvolume{96},
\bfpage{104}
(\byear{2017})
\end{barticle}
\endbibitem

\bibitem[\protect\citeauthoryear{Yuan et~al.}{2019}]{yuan2019examining}
\begin{barticle}
\bauthor{\bsnm{Yuan}, \binits{X.}},
\bauthor{\bsnm{Schuchard}, \binits{R.J.}},
\bauthor{\bsnm{Crooks}, \binits{A.T.}}:
\batitle{Examining emergent communities and social bots within the polarized online vaccination debate in twitter}.
\bjtitle{Social media+ society}
\bvolume{5}(\bissue{3}),
\bfpage{2056305119865465}
(\byear{2019})
\end{barticle}
\endbibitem

\bibitem[\protect\citeauthoryear{Cresci et~al.}{2017}]{cresci2017paradigm}
\begin{bchapter}
\bauthor{\bsnm{Cresci}, \binits{S.}},
\bauthor{\bsnm{Di~Pietro}, \binits{R.}},
\bauthor{\bsnm{Petrocchi}, \binits{M.}},
\bauthor{\bsnm{Spognardi}, \binits{A.}},
\bauthor{\bsnm{Tesconi}, \binits{M.}}:
\bctitle{The paradigm-shift of social spambots: Evidence, theories, and tools for the arms race}.
In: \bbtitle{Proceedings of the 26th International Conference on World Wide Web Companion},
pp. \bfpage{963}--\blpage{972}
(\byear{2017})
\end{bchapter}
\endbibitem

\bibitem[\protect\citeauthoryear{Ng and Carley}{2023}]{ng2023botbuster}
\begin{bchapter}
\bauthor{\bsnm{Ng}, \binits{L.H.X.}},
\bauthor{\bsnm{Carley}, \binits{K.M.}}:
\bctitle{Botbuster: Multi-platform bot detection using a mixture of experts}.
In: \bbtitle{Proceedings of the International AAAI Conference on Web and Social Media},
vol. \bseriesno{17},
pp. \bfpage{686}--\blpage{697}
(\byear{2023})
\end{bchapter}
\endbibitem

\bibitem[\protect\citeauthoryear{Shao et~al.}{2018}]{shao2018spread}
\begin{barticle}
\bauthor{\bsnm{Shao}, \binits{C.}},
\bauthor{\bsnm{Ciampaglia}, \binits{G.L.}},
\bauthor{\bsnm{Varol}, \binits{O.}},
\bauthor{\bsnm{Yang}, \binits{K.-C.}},
\bauthor{\bsnm{Flammini}, \binits{A.}},
\bauthor{\bsnm{Menczer}, \binits{F.}}:
\batitle{The spread of low-credibility content by social bots}.
\bjtitle{Nature communications}
\bvolume{9}(\bissue{1}),
\bfpage{1}--\blpage{9}
(\byear{2018})
\end{barticle}
\endbibitem

\bibitem[\protect\citeauthoryear{Geeng et~al.}{2020}]{geeng2020fake}
\begin{bchapter}
\bauthor{\bsnm{Geeng}, \binits{C.}},
\bauthor{\bsnm{Yee}, \binits{S.}},
\bauthor{\bsnm{Roesner}, \binits{F.}}:
\bctitle{Fake news on facebook and twitter: Investigating how people (don't) investigate}.
In: \bbtitle{Proceedings of the 2020 CHI Conference on Human Factors in Computing Systems},
pp. \bfpage{1}--\blpage{14}
(\byear{2020})
\end{bchapter}
\endbibitem

\bibitem[\protect\citeauthoryear{Wen et~al.}{2014}]{wen2014shut}
\begin{barticle}
\bauthor{\bsnm{Wen}, \binits{S.}},
\bauthor{\bsnm{Jiang}, \binits{J.}},
\bauthor{\bsnm{Xiang}, \binits{Y.}},
\bauthor{\bsnm{Yu}, \binits{S.}},
\bauthor{\bsnm{Zhou}, \binits{W.}},
\bauthor{\bsnm{Jia}, \binits{W.}}:
\batitle{To shut them up or to clarify: Restraining the spread of rumors in online social networks}.
\bjtitle{IEEE Transactions on Parallel and Distributed Systems}
\bvolume{25}(\bissue{12}),
\bfpage{3306}--\blpage{3316}
(\byear{2014})
\end{barticle}
\endbibitem

\bibitem[\protect\citeauthoryear{Alieva et~al.}{2022}]{alieva2022investigating}
\begin{bchapter}
\bauthor{\bsnm{Alieva}, \binits{I.}},
\bauthor{\bsnm{Ng}, \binits{L.H.X.}},
\bauthor{\bsnm{Carley}, \binits{K.M.}}:
\bctitle{Investigating the spread of russian disinformation about biolabs in ukraine on twitter using social network analysis}.
In: \bbtitle{2022 IEEE International Conference on Big Data (Big Data)},
pp. \bfpage{1770}--\blpage{1775}
(\byear{2022}).
\bcomment{IEEE}
\end{bchapter}
\endbibitem

\bibitem[\protect\citeauthoryear{Bragg et~al.}{2023}]{bragg2023less}
\begin{bchapter}
\bauthor{\bsnm{Bragg}, \binits{H.}},
\bauthor{\bsnm{Jayanetti}, \binits{H.R.}},
\bauthor{\bsnm{Nelson}, \binits{M.L.}},
\bauthor{\bsnm{Weigle}, \binits{M.C.}}:
\bctitle{Less than 4\% of archived instagram account pages for the disinformation dozen are replayable}.
In: \bbtitle{Proceedings of ACM/IEEE Joint Conference on Digital Libraries (JCDL)}
(\byear{2023})
\end{bchapter}
\endbibitem

\bibitem[\protect\citeauthoryear{Feng et~al.}{2022}]{feng2022twibot}
\begin{barticle}
\bauthor{\bsnm{Feng}, \binits{S.}},
\bauthor{\bsnm{Tan}, \binits{Z.}},
\bauthor{\bsnm{Wan}, \binits{H.}},
\bauthor{\bsnm{Wang}, \binits{N.}},
\bauthor{\bsnm{Chen}, \binits{Z.}},
\bauthor{\bsnm{Zhang}, \binits{B.}},
\bauthor{\bsnm{Zheng}, \binits{Q.}},
\bauthor{\bsnm{Zhang}, \binits{W.}},
\bauthor{\bsnm{Lei}, \binits{Z.}},
\bauthor{\bsnm{Yang}, \binits{S.}}, \betal:
\batitle{Twibot-22: Towards graph-based twitter bot detection}.
\bjtitle{Advances in Neural Information Processing Systems}
\bvolume{35},
\bfpage{35254}--\blpage{35269}
(\byear{2022})
\end{barticle}
\endbibitem

\bibitem[\protect\citeauthoryear{Heidari et~al.}{2021}]{heidari2021empirical}
\begin{bchapter}
\bauthor{\bsnm{Heidari}, \binits{M.}},
\bauthor{\bsnm{James~Jr}, \binits{H.}},
\bauthor{\bsnm{Uzuner}, \binits{O.}}:
\bctitle{An empirical study of machine learning algorithms for social media bot detection}.
In: \bbtitle{2021 IEEE International IOT, Electronics and Mechatronics Conference (IEMTRONICS)},
pp. \bfpage{1}--\blpage{5}
(\byear{2021}).
\bcomment{IEEE}
\end{bchapter}
\endbibitem

\bibitem[\protect\citeauthoryear{Chang et~al.}{2021}]{chang2021social}
\begin{botherref}
\oauthor{\bsnm{Chang}, \binits{H.-C.H.}},
\oauthor{\bsnm{Chen}, \binits{E.}},
\oauthor{\bsnm{Zhang}, \binits{M.}},
\oauthor{\bsnm{Muric}, \binits{G.}},
\oauthor{\bsnm{Ferrara}, \binits{E.}}:
Social bots and social media manipulation in 2020: the year in review.
arXiv preprint arXiv:2102.08436
(2021)
\end{botherref}
\endbibitem

\bibitem[\protect\citeauthoryear{Himelein-Wachowiak et~al.}{2021}]{himelein2021bots}
\begin{barticle}
\bauthor{\bsnm{Himelein-Wachowiak}, \binits{M.}},
\bauthor{\bsnm{Giorgi}, \binits{S.}},
\bauthor{\bsnm{Devoto}, \binits{A.}},
\bauthor{\bsnm{Rahman}, \binits{M.}},
\bauthor{\bsnm{Ungar}, \binits{L.}},
\bauthor{\bsnm{Schwartz}, \binits{H.A.}},
\bauthor{\bsnm{Epstein}, \binits{D.H.}},
\bauthor{\bsnm{Leggio}, \binits{L.}},
\bauthor{\bsnm{Curtis}, \binits{B.}}:
\batitle{Bots and misinformation spread on social media: Implications for covid-19}.
\bjtitle{Journal of medical Internet research}
\bvolume{23}(\bissue{5}),
\bfpage{26933}
(\byear{2021})
\end{barticle}
\endbibitem

\bibitem[\protect\citeauthoryear{Domashnev et~al.}{2019}]{domashnev2019usage}
\begin{barticle}
\bauthor{\bsnm{Domashnev}, \binits{P.}},
\bauthor{\bsnm{Alexeev}, \binits{V.}},
\bauthor{\bsnm{Lavrukhina}, \binits{T.}},
\bauthor{\bsnm{Nazarkin}, \binits{O.}}:
\batitle{Usage of telegram bots for message exchange in distributed computing}.
\bjtitle{International Journal of Open Information Technologies}
\bvolume{7}(\bissue{6}),
\bfpage{67}--\blpage{72}
(\byear{2019})
\end{barticle}
\endbibitem

\bibitem[\protect\citeauthoryear{de~Oliveira et~al.}{2016}]{de2016chatting}
\begin{bchapter}
\bauthor{\bsnm{Oliveira}, \binits{J.C.}},
\bauthor{\bsnm{Santos}, \binits{D.H.}},
\bauthor{\bsnm{Neto}, \binits{M.P.}}:
\bctitle{Chatting with arduino platform through telegram bot}.
In: \bbtitle{2016 IEEE International Symposium on Consumer Electronics (ISCE)},
pp. \bfpage{131}--\blpage{132}
(\byear{2016}).
\bcomment{IEEE}
\end{bchapter}
\endbibitem

\bibitem[\protect\citeauthoryear{Idhom et~al.}{2018}]{idhom2018implementation}
\begin{bchapter}
\bauthor{\bsnm{Idhom}, \binits{M.}},
\bauthor{\bsnm{Fauzi}, \binits{A.}},
\bauthor{\bsnm{Alit}, \binits{R.}},
\bauthor{\bsnm{Wahanani}, \binits{H.E.}}:
\bctitle{Implementation system telegram bot for monitoring linux server}.
In: \bbtitle{International Conference on Science and Technology (ICST 2018)},
pp. \bfpage{1089}--\blpage{1093}
(\byear{2018}).
\bcomment{Atlantis Press}
\end{bchapter}
\endbibitem

\bibitem[\protect\citeauthoryear{Alrhmoun et~al.}{2023}]{alrhmoun2023automating}
\begin{botherref}
\oauthor{\bsnm{Alrhmoun}, \binits{A.}},
\oauthor{\bsnm{Winter}, \binits{C.}},
\oauthor{\bsnm{Kert{\'e}sz}, \binits{J.}}:
Automating terror: The role and impact of telegram bots in the islamic state’s online ecosystem.
Terrorism and Political Violence,
1--16
(2023)
\end{botherref}
\endbibitem

\bibitem[\protect\citeauthoryear{Ng and Carley}{2022}]{ng2022pro}
\begin{barticle}
\bauthor{\bsnm{Ng}, \binits{L.H.X.}},
\bauthor{\bsnm{Carley}, \binits{K.M.}}:
\batitle{Pro or anti? a social influence model of online stance flipping}.
\bjtitle{IEEE Transactions on Network Science and Engineering}
\bvolume{10}(\bissue{1}),
\bfpage{3}--\blpage{19}
(\byear{2022})
\end{barticle}
\endbibitem

\bibitem[\protect\citeauthoryear{Hallgren}{2012}]{hallgren2012computing}
\begin{barticle}
\bauthor{\bsnm{Hallgren}, \binits{K.A.}}:
\batitle{Computing inter-rater reliability for observational data: an overview and tutorial}.
\bjtitle{Tutorials in quantitative methods for psychology}
\bvolume{8}(\bissue{1}),
\bfpage{23}
(\byear{2012})
\end{barticle}
\endbibitem

\bibitem[\protect\citeauthoryear{Artstein and Poesio}{2008}]{artstein2008inter}
\begin{barticle}
\bauthor{\bsnm{Artstein}, \binits{R.}},
\bauthor{\bsnm{Poesio}, \binits{M.}}:
\batitle{Inter-coder agreement for computational linguistics}.
\bjtitle{Computational linguistics}
\bvolume{34}(\bissue{4}),
\bfpage{555}--\blpage{596}
(\byear{2008})
\end{barticle}
\endbibitem

\bibitem[\protect\citeauthoryear{Grootendorst}{2022}]{grootendorst2022bertopic}
\begin{botherref}
\oauthor{\bsnm{Grootendorst}, \binits{M.}}:
Bertopic: Neural topic modeling with a class-based tf-idf procedure.
arXiv preprint arXiv:2203.05794
(2022)
\end{botherref}
\endbibitem

\bibitem[\protect\citeauthoryear{Tausczik and Pennebaker}{2010}]{tausczik2010psychological}
\begin{barticle}
\bauthor{\bsnm{Tausczik}, \binits{Y.R.}},
\bauthor{\bsnm{Pennebaker}, \binits{J.W.}}:
\batitle{The psychological meaning of words: Liwc and computerized text analysis methods}.
\bjtitle{Journal of language and social psychology}
\bvolume{29}(\bissue{1}),
\bfpage{24}--\blpage{54}
(\byear{2010})
\end{barticle}
\endbibitem

\bibitem[\protect\citeauthoryear{Kacewicz et~al.}{2014}]{kacewicz2014pronoun}
\begin{barticle}
\bauthor{\bsnm{Kacewicz}, \binits{E.}},
\bauthor{\bsnm{Pennebaker}, \binits{J.W.}},
\bauthor{\bsnm{Davis}, \binits{M.}},
\bauthor{\bsnm{Jeon}, \binits{M.}},
\bauthor{\bsnm{Graesser}, \binits{A.C.}}:
\batitle{Pronoun use reflects standings in social hierarchies}.
\bjtitle{Journal of Language and Social Psychology}
\bvolume{33}(\bissue{2}),
\bfpage{125}--\blpage{143}
(\byear{2014})
\end{barticle}
\endbibitem

\bibitem[\protect\citeauthoryear{Newman}{2005}]{newman2005measure}
\begin{barticle}
\bauthor{\bsnm{Newman}, \binits{M.E.}}:
\batitle{A measure of betweenness centrality based on random walks}.
\bjtitle{Social networks}
\bvolume{27}(\bissue{1}),
\bfpage{39}--\blpage{54}
(\byear{2005})
\end{barticle}
\endbibitem

\bibitem[\protect\citeauthoryear{Ng and Carley}{2023}]{ng2023combined}
\begin{barticle}
\bauthor{\bsnm{Ng}, \binits{L.H.X.}},
\bauthor{\bsnm{Carley}, \binits{K.M.}}:
\batitle{A combined synchronization index for evaluating collective action social media}.
\bjtitle{Applied network science}
\bvolume{8}(\bissue{1}),
\bfpage{1}
(\byear{2023})
\end{barticle}
\endbibitem

\bibitem[\protect\citeauthoryear{Cai et~al.}{2023}]{cai2023network}
\begin{barticle}
\bauthor{\bsnm{Cai}, \binits{M.}},
\bauthor{\bsnm{Luo}, \binits{H.}},
\bauthor{\bsnm{Meng}, \binits{X.}},
\bauthor{\bsnm{Cui}, \binits{Y.}},
\bauthor{\bsnm{Wang}, \binits{W.}}:
\batitle{Network distribution and sentiment interaction: Information diffusion mechanisms between social bots and human users on social media}.
\bjtitle{Information Processing \& Management}
\bvolume{60}(\bissue{2}),
\bfpage{103197}
(\byear{2023})
\end{barticle}
\endbibitem

\bibitem[\protect\citeauthoryear{Howard and Howard}{2019}]{howard2019bandwagon}
\begin{botherref}
\oauthor{\bsnm{Howard}, \binits{J.}},
\oauthor{\bsnm{Howard}, \binits{J.}}:
Bandwagon effect and authority bias.
Cognitive errors and diagnostic mistakes: A case-based guide to critical thinking in medicine,
21--56
(2019)
\end{botherref}
\endbibitem

\bibitem[\protect\citeauthoryear{Silvester}{2021}]{silvester2021authority}
\begin{botherref}
\oauthor{\bsnm{Silvester}, \binits{C.}}:
Authority bias.
Decision Making in Emergency Medicine: Biases, Errors and Solutions,
41--46
(2021)
\end{botherref}
\endbibitem

\bibitem[\protect\citeauthoryear{Duffy et~al.}{2020}]{duffy2020too}
\begin{barticle}
\bauthor{\bsnm{Duffy}, \binits{A.}},
\bauthor{\bsnm{Tandoc}, \binits{E.}},
\bauthor{\bsnm{Ling}, \binits{R.}}:
\batitle{Too good to be true, too good not to share: the social utility of fake news}.
\bjtitle{Information, Communication \& Society}
\bvolume{23}(\bissue{13}),
\bfpage{1965}--\blpage{1979}
(\byear{2020})
\end{barticle}
\endbibitem

\bibitem[\protect\citeauthoryear{Gilani et~al.}{2017}]{gilani2017bots}
\begin{bchapter}
\bauthor{\bsnm{Gilani}, \binits{Z.}},
\bauthor{\bsnm{Farahbakhsh}, \binits{R.}},
\bauthor{\bsnm{Tyson}, \binits{G.}},
\bauthor{\bsnm{Wang}, \binits{L.}},
\bauthor{\bsnm{Crowcroft}, \binits{J.}}:
\bctitle{Of bots and humans (on twitter)}.
In: \bbtitle{Proceedings of the 2017 IEEE/ACM International Conference on Advances in Social Networks Analysis and Mining 2017},
pp. \bfpage{349}--\blpage{354}
(\byear{2017})
\end{bchapter}
\endbibitem

\bibitem[\protect\citeauthoryear{Samper-Escalante et~al.}{2021}]{samper2021bot}
\begin{barticle}
\bauthor{\bsnm{Samper-Escalante}, \binits{L.D.}},
\bauthor{\bsnm{Loyola-Gonz{\'a}lez}, \binits{O.}},
\bauthor{\bsnm{Monroy}, \binits{R.}},
\bauthor{\bsnm{Medina-P{\'e}rez}, \binits{M.A.}}:
\batitle{Bot datasets on twitter: Analysis and challenges}.
\bjtitle{Applied Sciences}
\bvolume{11}(\bissue{9}),
\bfpage{4105}
(\byear{2021})
\end{barticle}
\endbibitem

\bibitem[\protect\citeauthoryear{Kloo and Carley}{2023}]{kloo2023social}
\begin{bchapter}
\bauthor{\bsnm{Kloo}, \binits{I.}},
\bauthor{\bsnm{Carley}, \binits{K.M.}}:
\bctitle{Social cybersecurity analysis of the telegram information environment during the 2022 invasion of ukraine}.
In: \bbtitle{International Conference on Social Computing, Behavioral-Cultural Modeling and Prediction and Behavior Representation in Modeling and Simulation},
pp. \bfpage{23}--\blpage{32}
(\byear{2023}).
\bcomment{Springer}
\end{bchapter}
\endbibitem

\end{thebibliography}

\end{document}